\documentclass[reprint,superscriptaddress,showkeys,amsmath,amssymb,aps]{revtex4-2}

\usepackage{graphicx}
\usepackage{dcolumn}
\usepackage{bm}
\usepackage{hyperref}
\usepackage{braket} 
\usepackage{fourier} 
\usepackage{upgreek} 
\usepackage[table,xcdraw]{xcolor} 
\usepackage{orcidlink}
\usepackage{lipsum} 

\usepackage{fancyhdr} 
\usepackage{titlesec}
\titleformat{\section}[runin]{\normalfont\bfseries\itshape}{\thesubsection}{0.5em}{}[\,\,\,--]
\titlespacing*{\section}{0pt}{*1}{5pt}

\begin{document}

\preprint{APS/123-QED}

\title{Electroweak Nuclear Properties from Single Molecular Ions in a Penning Trap}

\author{Karthein J.$^{1,*,\diamondsuit}$\orcidlink{0000-0002-4306-9708},
        Udrescu S.M.$^{1,*,\dagger}$\orcidlink{0000-0002-1989-576X},
        Moroch S.B.$^{1}$\orcidlink{0000-0002-3339-5083},
        Belosevic I.$^{2}$\orcidlink{0000-0002-8001-8889},
        Blaum K.$^{3}$\orcidlink{0000-0003-4468-9316},
        Borschevsky A.$^{4}$\orcidlink{0000-0002-6558-1921},\\
        Chamorro Y.$^{4}$\orcidlink{0000-0002-4664-2434},
        DeMille D.$^{5,6,\S}$\orcidlink{0000-0001-7139-4121},
        Dilling J.$^{7,8}$\orcidlink{0000-0002-8508-6426},
        Garcia Ruiz R.F.$^{1,\P}$\orcidlink{0000-0002-2926-5569},
        Hutzler N.R.$^{9}$\orcidlink{0000-0002-5203-3635},
        Pašteka L.F.$^{4,10}$\orcidlink{0000-0002-0617-0524},
        Ringle R.$^{11}$\orcidlink{0000-0002-7478-259X} \\~\\
\small \textit{$^{1}$Massachusetts Institute of Technology, Cambridge, MA 02139, USA}\\
\small \textit{$^{2}$TRIUMF, Vancouver, BC V6T 2A3, Canada}\\
\small \textit{$^{3}$Max Planck Institute for Nuclear Physics, 69117 Heidelberg, Germany}\\
\small \textit{$^{4}$University of Groningen, 9747AG Groningen, The Netherlands}\\
\small \textit{$^{5}$Department of Physics and James Franck Institute at the University of Chicago, Chicago, IL 60637, USA}\\
\small \textit{$^{6}$Physics Division at Argonne National Laboratory, Lemont, IL 60439, USA}\\
\small \textit{$^{7}$Duke University, Durham, NC 27708, USA}\\
\small \textit{$^{8}$Oak Ridge National Laboratory, Oak Ridge, TN 37830, USA}\\
\small \textit{$^{9}$California Institute of Technology, Pasadena, CA 91125, USA}\\
\small \textit{$^{10}$Comenius University, 84215 Bratislava, Slovakia}\\
\small \textit{$^{11}$Facility for Rare Isotope Beams, East Lansing, MI 48824, USA}\\~\\
\small \textit{Correspondence:} $^\diamondsuit$\href{mailto:karthein@mit.edu}{karthein@mit.edu}, $^\dagger$\href{mailto:sudrescu@mit.edu}{sudrescu@mit.edu}, $^\S$\href{mailto:ddemille@uchicago.edu}{ddemille@uchicago.edu}, $^\P$\href{mailto:rgarciar@mit.edu}{rgarciar@mit.edu}\\
\normalsize $^{*}$These authors contributed equally to this work.\\\vspace{5pt}}


\date{\today}

\begin{abstract}
\bf We present a novel technique to probe electroweak nuclear properties by measuring parity violation (PV) in single molecular ions in a Penning trap. The trap's strong magnetic field Zeeman shifts opposite-parity rotational and hyperfine molecular states into near degeneracy. The weak interaction-induced mixing between these degenerate states can be larger than in atoms by more than twelve orders of magnitude, thereby vastly amplifying PV effects. The single molecule sensitivity would be suitable for applications to nuclei across the nuclear chart, including rare and unstable nuclei.
\end{abstract}

\maketitle

\section{\label{sec:intro} Introduction}

Of Nature's four known fundamental forces, the weak force is the only one known to violate parity (P) and charge-parity (CP) symmetry. In this context, precision studies of the weak interaction provide powerful tests of the Standard Model (SM) \cite{Saf18}, violations of the fundamental symmetries, and the existence of new physics \cite{Dav12, Lan09, Saf18}. Accelerator-based experiments and atomic parity violation studies have provided key insights into the weak interaction between the electrons and nucleons, mediated by $Z^0$-boson exchange \cite{Woo97, PVDIS, QWEAK}. However, the electroweak interactions between nucleons are only poorly understood \cite{Des80, Ade85, Hax13, Gar17, deV20, Gar23}. A clear disagreement exists between measurements \cite{Ram06, Hol07, Saf18}.

Recent progress in precision control and interrogation of molecules has demonstrated powerful routes for precision studies of symmetry-violating properties \cite{ACM18, Cai17, Rou22, Saf18, Gar20}. Parity violation (PV) can produce unique signatures in the molecular energy levels, enabling the isolation of weak force effects from the overwhelmingly dominant strong and electromagnetic forces \cite{Sus78, DeM08, Fla85}. The proximity of opposite parity molecular levels provides high sensitivity to symmetry-violating properties, which can be several orders of magnitude larger than in atomic systems. Moreover, external magnetic fields can drive these opposite-parity states into near degeneracy, enhancing their sensitivity to PV properties \cite{Koz95}. The possibility of about eleven orders of magnitude of enhancement of PV-induced state mixing was recently demonstrated with a neutral beam of $^{138}$BaF \cite{Alt18}.

In this work, we propose and analyze a new method for measuring PV nuclear properties using single molecular ions and a Penning trap, which allows for long coherence times ($\gg1\,$ms) \cite{DeM08}. Combined with its well-controlled electric and magnetic fields, an enhancement in excess of twelve orders of magnitude in PV-induced state mixing relative to atoms can be achieved, thereby vastly increasing sensitivity to electroweak nuclear properties. The precision and versatility of our technique will enable measurements of many isotopes across the nuclear chart. These include species that may be difficult to manipulate and measure in neutral forms, such as short-lived nuclei \cite{Gar20, Udr21, RadMol2023}.

In a diatomic molecule, PV properties are dominated by the nuclear-spin-dependent interactions (NSD-PV): 

(i) Electrons penetrating the nucleus can interact at short range via $Z^0$-boson exchange through electron-vector and nucleon axial-vector currents \cite{Saf18};

(ii) Parity-violating weak interactions between nucleons lead to a nuclear-internal current that causes a P-odd magnetic moment, known as the nuclear anapole moment \cite{Fla80, Fla84, Saf18}. So far, only one non-zero measurement of the nuclear anapole moment has been performed in $^{133}$Cs \cite{Woo97};

(iii) A third contribution, typically suppressed compared to the effects above \cite{Dzu00, Joh03}, is induced by a combination of the hyperfine interaction and $Z^0$-boson exchange through electron-axial-vector and nucleon-vector $V_eA_N$ currents \cite{Saf18};

(iv) A fourth contribution could come from new interactions beyond the SM between electrons and nucleons, mediated by yet-to-be-discovered gauge bosons \cite{Dzu17, Sta15, Sta14}.

Our proposed method should be highly general for various molecular ions. However, we will focus on $^{29}$SiO$^+$ due to practical and theoretical advantages for the initial demonstration: Its rotational and electronic structure is known \cite{Sto16}, the ground electronic state is $^2\Sigma^+$, and it was demonstrated suitable for laser cooling \cite{Sto20, Ngu11}.

\section{\label{sec:eqs}Effective Hamiltonian and Electroweak Properties}

Our scheme builds on the concepts introduced in \cite{Koz91, DeM08}. The effective Hamiltonian describing the lowest rotational and hyperfine energy levels of $^{29}$SiO$^+$, in the absence of PV effects, can be expressed as:
\begin{equation*}
    H_0 = B_0\boldsymbol{N}^2+D_0\boldsymbol{N}^4+\gamma \boldsymbol{N}\cdot \boldsymbol{S} + b \boldsymbol{I}\cdot \boldsymbol{S}+c(\boldsymbol{I}\cdot \boldsymbol{n})(\boldsymbol{S}\cdot \boldsymbol{n}),
\end{equation*}
with $\boldsymbol{N} = \boldsymbol{R} + \boldsymbol{L}$, where $\boldsymbol{R}$ is the mechanical rotation of the molecular framework, $\boldsymbol{L}$ is the orbital angular momentum of the electron, $\boldsymbol{S}$ and $\boldsymbol{I}$ are the molecular frame electron and nuclear spin operator, respectively, and $\boldsymbol{n}$ is the unit vector along the internuclear axis. The rotational, centrifugal distortion, and spin-rotational constants are $B_0$, $D_0$, and $\gamma$. $b$ and $c$ are hyperfine structure constants associated with the $^{29}$Si nucleus. The rotational constant of $^{29}$SiO$^+$ is far larger than all the other molecular parameters in $H_0$ \cite{Zhu22}. Thus, $N$ is a good quantum number for levels of energy $E_N \approx B_0N(N+1)$ and parity $P_N = (-1)^N$.

\begin{figure}[t!]
    \centering
    \includegraphics[width=\linewidth]{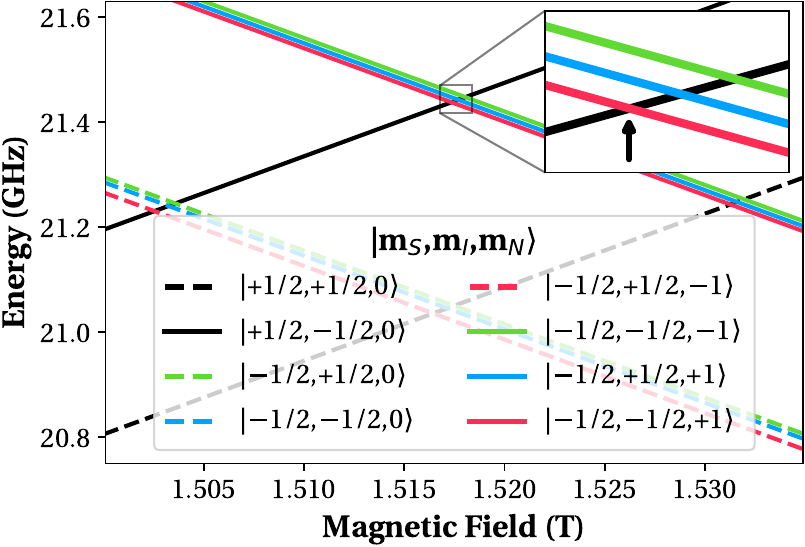}
    \caption{Calculated energies of opposite parity rotational and hyperfine states in $^{29}$SiO$^+$ for different magnetic field strengths, based on the Hamiltonian $H_0$ and parameters given in \cite{knight1985generation,zhu2022high}. Near degeneracy can be achieved at $B\approx1.5\,$T, and one possible crossing useful for detecting PV is indicated by an arrow. The positive parity states $|\varPsi_\uparrow^+\rangle$ are rising, while the negative ones $|\varPsi_\downarrow^-\rangle$ are descending.}
    \label{fig:1-SiO-state-crossing}
\end{figure}

When a magnetic field of a particular magnitude $B$ is applied (see Fig.\,\ref{fig:1-SiO-state-crossing}), sub-levels of the $N^P = 0^+$ and $1^-$ states can be Zeeman-shifted close to degeneracy. For $^{29}$SiO$^+$, this magnetic field strength is $B\approx\frac{E_1-E_0}{2\mu_B} \approx 1.5\,$T, since the coupling to the electron spin $S$ dominates the Zeeman shift via the Hamiltonian $H_Z = -g\mu_B\boldsymbol{S}\cdot \boldsymbol{B}$ with $g$-factor $g\approx-2$, the Bohr magneton $\mu_B$, and the magnetic field aligned with the $\boldsymbol{z}$-axis $\boldsymbol{B}=B\boldsymbol{z}$ \cite{Alt18}. This field is strong enough to decouple $\boldsymbol{S}$ from $\boldsymbol{I}$ and $\boldsymbol{N}$. Hence, the rotational and hyperfine levels are better described in the decoupled basis used for the rest of the paper: $|N,m_N\rangle|S,m_S\rangle|I,m_I\rangle$.

The NSD-PV interactions can mix opposite-parity levels. The Hamiltonian $H_{\rm{PV}} =\kappa'\frac{G_F}{\sqrt{2}}\frac{\boldsymbol{\alpha} \boldsymbol{I}}{I}\rho(\boldsymbol{r})$ \cite{Fla80} describes such PV interactions, where $\kappa'$ includes all the NSP-PV contributions. We denote the Fermi constant $G_F$, Dirac matrices vector $\boldsymbol{\alpha}$, nuclear spin $\boldsymbol{I}$, and nuclear density with respect to the nuclear center $\rho(\boldsymbol{r})$. An effective Hamiltonian acting only within the subspace of rotational and hyperfine levels can be obtained by averaging the previous Hamiltonian over the electronic wave function, given by $H_{\rm{eff}} = \kappa'W_{\rm{A}}C$, where $W_{\rm{A}}$ is a matrix element that includes the expectation value of $H_{\rm{PV}}$ over the electronic wave function in the $^2\Sigma$-state in the rotating frame of the molecule, which can be computed numerically using state-of-the-art quantum chemistry methods with uncertainties as low as a few percent \cite{Hao18}. $C=\frac{(\boldsymbol{n}\times \boldsymbol{S})\cdot \boldsymbol{I}}{I}$ contains the angular momentum dependence of $H_{\rm{eff}}$ and its matrix elements can be calculated analytically using angular momentum algebra \cite{Fla85}. 

\section{\label{sec:meas}Measurement Strategy}

Our proposed experiment will be performed in a Penning ion trap. This device is widely used in precision atomic and nuclear physics, providing the highest mass accuracy \cite{Fil21} and longest trapping times of stable \cite{Sai22}, radioactive \cite{Mou21}, and antimatter particles \cite{Bor22}. The trap consists of a strong magnetic and a weak electrostatic field, allowing three-dimensional trapping of ions (see \cite{Bro86} for a review on Penning traps). We take advantage of the trapping magnetic field to Zeeman-shift two opposite parity states into near degeneracy (see arrow in Fig.\,\ref{fig:1-SiO-state-crossing}). Moreover, the intrinsic trap design allows for magnetic field strengths up to $12\,$T \cite{Mar14}, thus providing maximal flexibility in the choice of ion species and rotational-hyperfine states.

Our experimental principle is identical to the one for neutral molecules in Refs. \cite{DeM08, Alt18}. In the presence of axial (i.e., aligned with the magnetic field) and radial electric fields, $E_z$ and $E_r$, the effective Hamiltonian of this two-level system is: 
\begin{equation*}
    H_\pm =
    \begin{pmatrix}
        \alpha_rE_r^2 + \alpha_zE_z^2 & iW+d\cdot E_z  \\
        -iW+d\cdot E_z & \varDelta
    \end{pmatrix} ,
\end{equation*}
with the weak interaction matrix element
\begin{equation*}
\begin{split}
    iW(m'_N, m'_I, &m_N, m_I) \equiv \\ & \kappa'W_{\rm{A}}\langle \mathit{\varPsi}_\downarrow^-(m'_N, m'_I)|C|\mathit{\varPsi}_\uparrow^+(m_N, m_I)\rangle,
\end{split}
\end{equation*}
the expectation value $d$ of the dipole moment operator, $\boldsymbol{D}$, between the two levels and the general wave function, $|\varPsi(t)\rangle= c_+(t)|\varPsi_\uparrow^+\rangle+e^{-i\varDelta t}c_-(t)|\varPsi_\downarrow^-\rangle$, of the two-level system with its eigenstates $|\varPsi^P_{m_S}\rangle$ of parity $P$ and spin projection $m_S$, and its time-dependent amplitudes $c_P(t)$ (see Refs. \cite{DeM08,Alt18} and the Supplemental Material (SM)-\ref{sec:analyt-formula} for details).
$\varDelta$ is a small detuning of the two levels from perfect degeneracy and depends on the applied magnetic field strength $B$; $\alpha_r$ and $\alpha_z$ represent the radial and axial contributions to the differential polarizability of the two levels \cite{Cah14}, while $E_r$ and $E_z$ are any external radial and axial $E$-fields.

In the ideal case of a single ion resting in a stable magnetic field $B$ at $t_0=0$ with zero external electric fields prepared in the $|\varPsi_\uparrow^+\rangle$ state in the center of our trap, we measure $W$ using the Stark-interference procedure described in Ref. \cite{DeM08}. Thereby, we "kick" the ion to a well-defined amplitude in the harmonic trapping potential, leading to an electric field $E_z(t) = E_{\rm{ext}}\cdot \rm{sin}(\omega_{\rm{ext}}\it{t})$ experienced in the ion's rest frame. We repeat this measurement for several $N_{\rm{0}}$ ions to determine the population transfer probability from the initial to the other parity state, $|\varPsi_\downarrow^-\rangle$, by measuring the average signal $S = N_0|c_-(t)|^2$ (see SM-\ref{sec:analyt-formula} for details). The existence of parity violation leads to a non-zero asymmetry, defined as $A_{\rm{PV}} \equiv \frac{S(+E_{\rm{ext}})-S(-E_{\rm{ext}})}{S(+E_{\rm{ext}})+S(-E_{\rm{ext}})}$ \cite{Alt18}, where $S(+E_{\rm{ext}})$ and $S(-E_{\rm{ext}})$ refer to the signals obtained for measurements with the initial "kick" applied in positive (+) or negative ($-$) axial direction.

For $^{29}$SiO$^+$, the population transfer and, hence, the asymmetry can be estimated using first-order perturbation theory (see SM-\ref{sec:analyt-formula} for details). For interrogation times $t_{\rm{x}}\approx\frac{2\pi N}{\omega_{\rm{ext}}}\approx\frac{\pi}{\varDelta}$ at integer $N$, the PV asymmetry becomes \cite{DeM08}:
\begin{equation}\label{eq:A_PV}
    A_{\rm{PV}} = \frac{\frac{2W}{\varDelta}\cdot\frac{\varOmega_{\rm{R}}}{\omega_{\rm{ext}}}}{\left(\frac{W}{\varDelta}\right)^2+\left(\frac{\varOmega_{\rm{R}}}{\omega_{\rm{ext}}}\right)^2},
\end{equation}
with $\varOmega_{\rm{R}}=dE_{\rm{ext}}$. Ultimately, $W$ is determined via the population transfer probability for different values of $\varDelta$, i.e., magnetic field strengths $B$ we can easily scan in our setup. Its statistical uncertainty is
\begin{equation}
    \updelta W = \frac{\varDelta}{4\sqrt{2N_0}\sin\left(\frac{\varDelta t_{\rm{x}}}{2}\right)}\frac{\sqrt{\eta^2+1}}{\eta}
\end{equation} using $\eta \equiv \left(\frac{\Omega_R}{\omega_{\mathrm{ext}}}\right)/\left(\frac{W}{\varDelta}\right)$ for the number of molecules $N_0$.

To reduce $\updelta W$, we want to minimize $\varDelta$. Since we are technically limited in arbitrarily reducing $\varDelta$ (as discussed in the following section), we set the interrogation time to $t_{\rm{x}}=\frac{\pi}{\varDelta}$ once $\varDelta$ is minimized. Thus, the precise control of the interrogation time $t_{\rm{x}}$ in our trap for a minimal uncertainty on $\updelta W$ and precise variation of $t_{\rm{x}}$ to check for systematic effects, are clear advantages we can leverage over experiments performed on molecular beams.

From our measurement of $W$ and the calculated $W_{\rm{A}}$ and $C$, we can extract $\kappa' \approx \kappa'_2 + \kappa'_{\rm{a}}$, encoding the physics of the weak interaction that leads to NSD-PV: $\kappa'_2$, arising from the $V_eA_N$ term in the electron-nucleon-$Z^0$-boson exchange, and the electron electromagnetic interaction with the anapole moment, $\kappa'_{\rm{a}}$. Applying our technique to a wide range of isotopic chains, including radioactive ones \cite{Gar20, Udr21, RadMol2023}, could possibly allow for a separation of $\kappa'_2$ and $\kappa'_{\rm{a}}$ based on the dependence of $\kappa'_{\rm{a}}$ on the nuclear mass $A$ and spin $I$ \cite{Fla84, DeM08}.

\section{Experimental Details}

Trapped ions in a Penning trap move on three superimposed eigenmotions inside the trap: two radial ones perpendicular to the magnetic field and one axial along the magnetic field. The eigenmotions' frequency, phase, and amplitude can be controlled and coupled through radio-frequency excitations on the ion trap's electrodes \cite{Bro86}. The eigenmotions can be further cooled by coupling the axial motion to a resonance circuit at $1\,$K. The radial eigenmotions can be cooled to the same temperature by side-band coupling to the axial eigenmotion \cite{Cor90}. Once the ion is located in the trap center in equilibrium with the 1-K-environment, it is decoupled from the resonance circuit using a cryogenic switch. It remains in a nominally zero $E_{\rm{ext}}$-field, allowing for the above assumptions on the Hamiltonian due to low reheating rates of $\sim65\,$mK/s \cite{Jen04}.

An additional, significant advantage of our proposed method is that the magnetic field strength $B$ experienced by the molecular ion with charge-to-mass ratio $q/m$ can be precisely determined through a cyclotron frequency $\nu_c=\frac{Bq}{2\pi m}$ determination via the Fourier-transform ion-cyclotron-resonance (FT-ICR) method \cite{Com74} to the $10^{-11}$ level of precision \cite{Sch20} or better using a cryogenic resonance circuit of high quality ($Q>5000$).

\begin{figure}[t!]
    \centering
    \includegraphics[width=\linewidth]{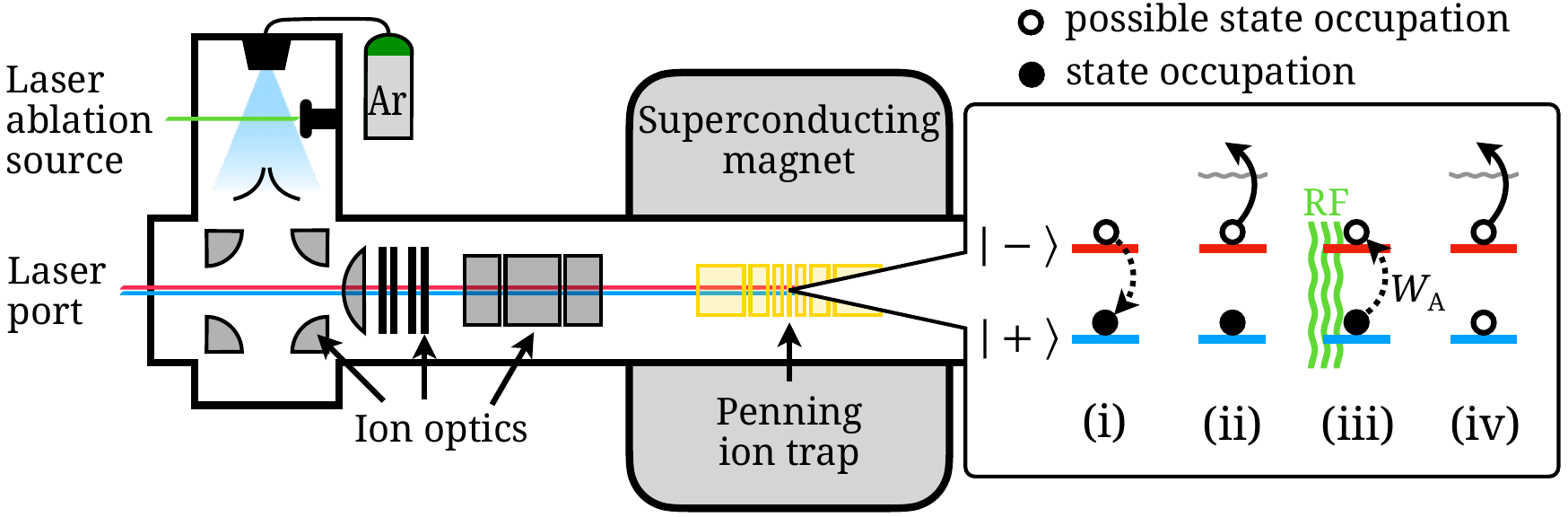}
    \caption{Schematic layout and measurement principle with a laser port for the ionization, cooling, and dissociation lasers. Our measurement procedure, (i)-(iv), is described in the text.}
    \label{fig:2-scheme}
\end{figure}

In our proposed setup, neutral $^{29}$SiO molecules are produced by laser-ablating a silicon rod in the supersonic expansion of a mixture of oxygen and argon gas \cite{Mar96}. The molecules are photo-ionized using resonant laser light \cite{Sto19} and bend towards the Penning trap. The ions are produced into the ground electronic and vibrational states and populate only low rotational levels \cite{Ton10}. The measurement scheme shown in Fig.\,\ref{fig:2-scheme} works as follows:

(i) The molecular ions are trapped in the Penning trap, and a single molecule is selected using the evaporative cooling technique \cite{And10}. Once the ion is located at the trap center in equilibrium with the 1-K-environment (assumed as the kinetic temperature of the ions moving forward) and decoupled from the resonant circuit, it is optically pumped into its rotational ground state (94(3)\% fidelity were shown in Ref. \cite{Sto20} for $^{28}$SiO$^+$). This level is further split into four hyperfine substates. Given the large splitting between these substates ($>100\,$MHz), they can be addressed individually after the rotational cooling using lasers or microwaves to transfer the population to the state of interest, $|\varPsi_\uparrow^+\rangle$ (Fig.\,\ref{fig:1-SiO-state-crossing}, solid black line), with $>90\%$ fidelity.

(ii) To ensure the molecule is not in the negative parity state $|\varPsi_\downarrow^-\rangle$ (Fig.\,\ref{fig:1-SiO-state-crossing}, colored lines) even after the state transfer, the molecule in $|\varPsi_\downarrow^-\rangle$ is state-selectively dissociated via excitation to a higher-lying auto-dissociating state \cite{Sto20}. The time scale for this process is $\sim10\,$ns, i.e., short compared to all inverse frequencies in this measurement; thus, it corresponds to an instantaneous (but conditional) quantum projection onto unaffected states.

(iii) This step constitutes the starting point of the measurement. It will be executed after step (i) and in parallel to step (ii) since $|\varPsi_\uparrow^+\rangle$ would start to evolve in time even without an external electric field.

In the ion's rest frame, we have it experience a sinusoidal electric field $E_z(t) = E_{\rm{ext}}\cdot \rm{sin}(\omega_{\rm{ext}} \it t)$ with $E_{\rm{ext}}\approx 6\,$V/cm and $\varOmega_{\rm{R}}/2\pi \approx 3\,$kHz. This is achieved by exciting the ion to an axial amplitude of $\sim 0.3\,$mm in the harmonic trapping potential with a $\sim20\,$V single cycle, resonant sinusoidal-wave "kick" to the trap's end caps as routinely achieved in practice \cite{Stu19}. Population transfer from the initial positive to the negative parity state occurs due to the PV matrix element and the interaction with this sinusoidal electric field.

The minimum useful working value of the splitting is limited by the uncertainty associated with $\varDelta$. The main contribution to this effect is expected to come from the AC Stark shift of the energy levels of interest due to the transverse and axial components of the electric field, with the effects proportional to $\alpha_rE_r^2$ and $\alpha_zE_z^2$, respectively. The uncertainty associated with this shift arising from the thermal distribution of ion positions and velocities is expected to be $\updelta\varDelta/2\pi \approx30\,$Hz (see SM-\ref{sec:rad-E-field} for details of the calculations). To clearly tell apart the two opposite parity levels of interest, we assume a value of $\varDelta/2\pi \approx 100\,$Hz, and therefore $t_{\rm{x}} = \pi/\varDelta = 5\,$ms to minimize $\updelta W$.

(iv) The final state detection is performed by molecular dissociation of the negative parity state $|\varPsi_\downarrow^-\rangle$, using the same auto-ionizing state as in step (ii) as soon as the oscillating field in step (iii) is switched "off" by reversing the sinusoidal "kick". Since the dissociation process is parity-state selective, we can perform a "double-dip" mass measurement \cite{Stu19} in search of $^{29}$SiO$^+$, $^{29}$Si$^+$, or $^{16}$O$^+$ as a measurement of the final parity state. If a dissociation had occurred, we can remove the $^{29}$Si$^+$ or $^{16}$O$^+$ ion from the trap and load a new $^{29}$SiO$^+$ ion. If no dissociation had occurred, the measurement would be restarted at step (i).

Figure \ref{fig:3-pop-asym} shows the simulated PV asymmetry, $A_{PV}$, in Eq. \ref{eq:A_PV}, as a function of $\varDelta$ for a range of possible $W$ values. For $^{29}$SiO$^+$, we assume $\varOmega_{\rm{R}}/2\pi = 3\,$kHz, $\omega_{\rm{ext}}/2\pi = 350 \,$kHz, and scan $\varDelta/2\pi$ ranging from $-150\,$Hz to $150\,$Hz in steps of $50\,$Hz. Measuring different values of $\varDelta$ was shown to be effective in avoiding various systematic uncertainties \cite{Alt18, Alt18b}. Measuring also at other relevant level crossings will allow diagnosing systematics.

Heavier molecules with larger weak matrix elements comparable to $\varDelta$ ($W\gtrsim100\,$Hz), such as the potentially laser-coolable TlF$^+$ \cite{Chm21} (see Tab. \ref{tab:W_A}), do not require additional external Stark mixing for amplifying the sought signal. As suggested in Ref. \cite{Koz91}, the level crossing shown in Fig. \ref{fig:1-SiO-state-crossing} turns into a pseudo crossing, which can be measured directly. This approach requires an advanced level of systematic control planned to be investigated in the future.

\section{Uncertainty Estimates}\label{sec:unc}

\begin{figure}[t!]
    \centering
    \includegraphics[width=\linewidth]{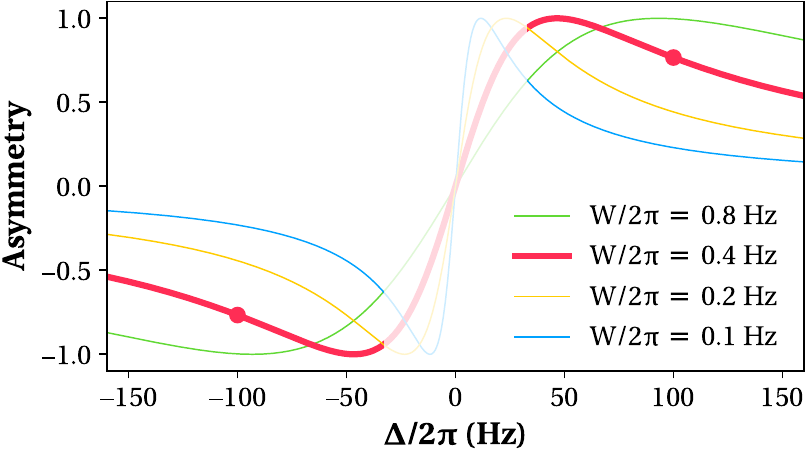}
    \caption{Asymmetry for $^{29}$SiO$^+$ (with external electric field; $\varOmega_{\rm{R}}/2\pi = 3\,$kHz, $\omega_{\rm{ext}}/2\pi = 350\,$kHz) for different $W$ and $\varDelta$. The assumed experimental condition is indicated in red for the calculated $W_{\rm{A}}/2\pi=16\,$Hz from Tab. \ref{tab:W_A}, corresponding to $W/2\pi = 0.4\,$Hz when assuming $\kappa'=0.05$ and $C=0.5$. The red dots show the expected asymmetry at $\varDelta/2\pi = \pm100\,$Hz}
    \label{fig:3-pop-asym}
\end{figure}

Here, we estimate the primary sources and magnitude of uncertainty for $^{29}$SiO$^+$ with $\varDelta/2\pi = 100\,$Hz and $W/2\pi = 0.4\,$Hz. These values lead to a maximum state transfer probability of the positive parity state's population of $\sim0.06\%$ after $t_{\rm{x}} = 5\,$ms for $^{29}$SiO$^+$, corresponding to an asymmetry of $\sim 0.75$ (red dots in Fig. \ref{fig:3-pop-asym}). 

(i) \textit{Initial Axial Amplitude}$\,\,\,$--$\,\,\,$Besides the already mentioned induced AC Stark shift of the energy levels of interest, leading to $\updelta\varDelta/2\pi \approx 30\,$Hz, i.e., $\updelta W/W\approx 30$\% from a single observed state transfer event, a second major source of uncertainty is expected to derive from the thermal noise in the initial axial amplitude of cooled ions. Once cooled and resting in the center of the Penning trap, the ions' energy is Boltzmann distributed with an average initial axial amplitude of $z_0=\sqrt{\frac{2k_{\rm{b}}Td^2_{\rm{char}}}{q_{\rm{ion}}U_0C_2}}$, where $k_{\rm{b}}$ is the Boltzmann constant, $q_{\rm{ion}}$ is the electron charge $e$, and we assume $T = 1\,$K. Based on our trap design \cite{Stu19} optimized for $E$-field homogeneity of the electric quadrupole potential $\phi(z,\rho)=\frac{|U_0|C_2}{2d_{\rm{char}}}\left(z^2-\rho^2/2\right)$, we further assumed for the characteristic trap length $d_{\rm{char}}=\sqrt{0.5(z_{\rm{trap}}^2+r_{\rm{trap}}^2/2)}= 3\,$mm (with the central ring electrode's length $z_{\rm{trap}}$ and radius $r_{\rm{trap}}$), the trap potential $U_0 = -85\,$V, and the dimensionless quadrupole constant $C_2 = -0.6$. The initial axial motion is then $z_0 \approx 10\,\upmu$m, which would result in an average thermal noise of $\updelta E_{\rm{th}} \approx 0.2\,$V/cm, corresponding to $\updelta W/W \approx 3$\% for $^{29}$SiO$^+$. Both of these effects are statistical, i.e., they can be reduced by increasing the number of measurements.

(ii) \textit{Magnetic Field}$\,\,\,$--$\,\,\,$Short-term magnetic field instabilities (for the measurement time of up to many milliseconds) are expected to be $\updelta B/B \lesssim 10^{-10}$ \cite{Tak22, Bri16}. Observed temporal changes in the magnetic field tracked in a neighboring trap center will be used for live adjustment of slow magnetic field drifts on top of typical temperature and pressure stabilization of the magnet \cite{Stu19}. With this method we anticipate $\updelta B/B \approx 10^{-10}$ for the duration of the data taking~\cite{Dro11}. Furthermore, deviations from spatial uniformity due to higher-order field effects not accounted for by shimming coils are expected to be $\updelta B/B < 10^{-10}$ for the small probed volume of $\ll0.1\,$mm$^3$ \cite{Stu19}. All of these effects can be quantified based on precise measurements of $\nu_{\rm{c}}$ for well-known species. These effects lead to a total systematic uncertainty from the magnetic field of $\updelta B/B \approx 10^{-10}$, or $\updelta\varDelta/2\pi \approx 4\,$Hz, i.e., $\updelta W/W \approx 4$\% for $^{29}$SiO$^+$. This uncertainty can be reduced by at least one order of magnitude by improving the stability and uniformity of the magnetic field.

(iii) \textit{Electric Field}$\,\,\,$--$\,\,\,$A relative electric field uncertainty of $\updelta E/E \ll 1\%$, which can be routinely achieved in practice \cite{Stu19}, would have negligible effect on $\updelta W$.

We thus anticipate a total systematic uncertainty of $\updelta W/W < 5$\% for $^{29}$SiO$^+$. To achieve a $10\%$ statistical uncertainty on the proposed measurement, we need on the order of $10^5$ trapped molecular ions. Given a measurement cycle of a few seconds (dominated by mass selection, cooling, and state preparation), a $10\%$ relative uncertainty measurement would thus be feasible in about one week of measurement time for $^{29}$SiO$^+$.

\section{Calculated Sensitivity Factors}

We calculated the molecular matrix element of the anapole moment $W_{\rm{A}}$ for the $^2\Sigma_{1/2}$ ground states of BF$^+$, $^{29}$SiO$^+$, and TlF$^+$ at the 4-component relativistic Fock-space coupled-cluster (FSCC) level of theory using the finite field approach. This formalism includes $H_{\rm{PV}}$ as a perturbation to the Dirac–Coulomb Hamiltonian. The $W_{\rm{A}}$ factor is obtained as the first derivative of the total energy to this perturbation \cite{Hao18}. We used the dyall.cv4z basis sets \cite{Dya06, Dya16} and correlated 13 (all), 21 (all), and 51 electrons for BF$^+$, $^{29}$SiO$^+$, and TlF$^+$, respectively. A Gaussian charge distribution represented the nucleus. All the calculations were performed using an adapted version of the Dirac program package \cite{Gom19, Sau20}.

Furthermore, we calculated $W_{\rm{A}}$ for Ac, Th, and Lr-containing molecular ions. Here, we used the 4-component relativistic Dirac–Hartree–Fock (DHF) level of theory. In this case, $W_{\rm{A}}$ was extracted from the off-diagonal matrix elements of the operator $\alpha\rho(r)$ acting on the degenerate $\varOmega =|\pm1/2\rangle$ states in the molecular spinor basis. We employed the dyall.cv4z basis set for all the elements \cite{Dya02, Dya06, Dya07, Dya16}.

The molecular geometries were optimized at the exact 2-component \cite{Ili07, Sau11} coupled-cluster level of theory, including single and double excitations in the parallel implementation of the Dirac program package \cite{Pot21}. The cut-off was set to $-20$ to 30 a.u. We used the dyall.v3z basis sets \cite{Dya02, Dya07, Dya16} for all the systems, except for $^{29}$SiO$^+$ (experimental bond length \cite{Lag73}), and BF$^+$/TlF$^+$ (s-aug-dyall.v4z basis sets \cite{Dya06, Dya16}). All results are presented in Table~\ref{tab:W_A}.

Besides $^{29}$SiO$^+$ \cite{Sto16, Sto20, Ngu11}, spectroscopic information in the literature among the presented molecular ions is not available to the best of our knowledge. Hence, prior studies of each molecular ion are necessary to find the needed rotational/hyperfine parameters and laser-cooling transitions.

\section{Outlook}

We proposed a new technique that can provide a highly sensitive route to investigate yet-to-be-explored nuclear parity-violating properties using single molecular ions. These measurements will enable stringent tests of the weak interaction in stable and short-lived isotopes across the nuclear chart. This technique could be directly applied to light isotopes, for which PV nuclear properties can already be calculated on the lattice \cite{Kur16, Dav21} and with ab-inito methods \cite{Hao20}. For diatomic molecules containing elements as light as the deuteron, the magnetic fields for ground-state level-crossings exceed the latest magnet technology in diatomic molecules; however, this challenge could be overcome by using ground-rotational states in polyatomic molecules \cite{DeM08, Nor19}. Furthermore, applying advanced cooling techniques already demonstrated in Penning traps would enable reducing the trapped molecule's kinetic energy even further to $\sim10-100\,$mK \cite{Boh21, Wil22, Wil23} or even $\sim1\,$mK \cite{Ita82, Tor16}, resulting in a reduction of the uncertainty on $W$ by one to two orders of magnitude.

\begin{table}[t!]
\caption{Diatomic molecular ions with sizable weak matrix elements $W_{\rm{A}}$ (of the first-mentioned atom) in units of Hz, calculated using the FSCC method for BF$^+$, $^{29}$SiO$^+$, and TlF$^+$, the DHF method for molecules containing Ac/Th/Lr. Data on additional molecular ions can be found in \cite{Bor12}.}
\renewcommand{\arraystretch}{1.2}
\begin{tabular}{|cc||cc||cc|}
\hline
\textbf{System}                            & \textbf{$\boldsymbol{W}_{\rm{\textbf{A}}}$}            & \textbf{System} & \textbf{$\boldsymbol{W}_{\rm{\textbf{A}}}$} & \textbf{System} & \textbf{$\boldsymbol{W}_{\rm{\textbf{A}}}$} \\
\hline
 $^{11}$B$^{19}$F$^+$  & 1    & $^{227}$Ac$^{79}$Br$^+$      & 2050                      & $^{261}$Lr$^{1}$H$^+$        & 14088                     \\
$^{29}$Si$^{16}$O$^+$  & 16   & $^{227}$Ac$^{127}$I$^+$      & 2092                      & $^{261}$Lr$^{7}$Li$^+$       & 3424                      \\
$^{205}$Tl$^{19}$F$^+$ & 5578                                 & $^{229}$Th$^{16}$O+       & 3126                      & $^{261}$Lr$^{19}$F$^+$       & 11009                     \\
$^{227}$Ac$^{1}$H$^+$                                   & 2057                                  & $^{229}$Th$^{32}$S$^+$       & 2186                      & $^{261}$Lr$^{23}$Na$^+$      & 3025                      \\
$^{227}$Ac$^{19}$F$^+$                                  & 2065                                  & $^{229}$Th$^{80}$Se$^+$      & 1981                      & $^{261}$Lr$^{35}$Cl$^+$      & 12671                     \\
$^{227}$Ac$^{35}$Cl$^+$                                 & 2035                                  & $^{229}$Th$^{126}$Te$^+$     & 1650                      & $^{261}$Lr$^{39}$K$^+$       & 2069                     \\
\hline
\end{tabular}\label{tab:W_A}
\end{table}

\begin{acknowledgments}
This work was supported by the U.S. Department of Energy (DOE), Office of Science (OS), and Office of Nuclear Physics under Award numbers DE-SC0021176 and DE-SC0021179. This research is partly based on work supported by Laboratory Directed Research and Development (LDRD) funding from Argonne National Laboratory, provided by the OS Director of the U.S. DOE under Contract DE-AC02-06CH11357. We thank the Center for Information Technology of the University of Groningen for its support and access to the Peregrine high-performance computing cluster. The INCITE program awarded computer time. This research also used resources from the Oak Ridge Leadership Computing Facility, a DOE-OS User Facility supported under Contract DE-AC05-00OR22725. We also acknowledge the support from High Sector Fock space coupled cluster method: benchmark accuracy across the periodic table (with project number VI.Vidi.192.088 of the research program Vidi, financed by the Dutch Research Council) and the 2020 Incite Award: “PRECISE: Predictive Electronic Structure Modeling of Heavy Elements.” JK acknowledges the support of a Feodor Lynen Fellowship of the Alexander-von-Humboldt Foundation. SBM acknowledges the support of a National Science Foundation Graduate Research Fellowship (NSF Grant \#2141064) and a Fannie and John Hertz Graduate Fellowship.
\end{acknowledgments}

\bibliography{PenningTrapPV}

\begin{thebibliography}{84}%
\makeatletter
\providecommand \@ifxundefined [1]{%
 \@ifx{#1\undefined}
}%
\providecommand \@ifnum [1]{%
 \ifnum #1\expandafter \@firstoftwo
 \else \expandafter \@secondoftwo
 \fi
}%
\providecommand \@ifx [1]{%
 \ifx #1\expandafter \@firstoftwo
 \else \expandafter \@secondoftwo
 \fi
}%
\providecommand \natexlab [1]{#1}%
\providecommand \enquote  [1]{``#1''}%
\providecommand \bibnamefont  [1]{#1}%
\providecommand \bibfnamefont [1]{#1}%
\providecommand \citenamefont [1]{#1}%
\providecommand \href@noop [0]{\@secondoftwo}%
\providecommand \href [0]{\begingroup \@sanitize@url \@href}%
\providecommand \@href[1]{\@@startlink{#1}\@@href}%
\providecommand \@@href[1]{\endgroup#1\@@endlink}%
\providecommand \@sanitize@url [0]{\catcode `\\12\catcode `\$12\catcode
  `\&12\catcode `\#12\catcode `\^12\catcode `\_12\catcode `\%12\relax}%
\providecommand \@@startlink[1]{}%
\providecommand \@@endlink[0]{}%
\providecommand \url  [0]{\begingroup\@sanitize@url \@url }%
\providecommand \@url [1]{\endgroup\@href {#1}{\urlprefix }}%
\providecommand \urlprefix  [0]{URL }%
\providecommand \Eprint [0]{\href }%
\providecommand \doibase [0]{https://doi.org/}%
\providecommand \selectlanguage [0]{\@gobble}%
\providecommand \bibinfo  [0]{\@secondoftwo}%
\providecommand \bibfield  [0]{\@secondoftwo}%
\providecommand \translation [1]{[#1]}%
\providecommand \BibitemOpen [0]{}%
\providecommand \bibitemStop [0]{}%
\providecommand \bibitemNoStop [0]{.\EOS\space}%
\providecommand \EOS [0]{\spacefactor3000\relax}%
\providecommand \BibitemShut  [1]{\csname bibitem#1\endcsname}%
\let\auto@bib@innerbib\@empty
\bibitem [{\citenamefont {{Safranova M.S. \it{et al.}}}(2018)}]{Saf18}%
  \BibitemOpen
  \bibfield  {author} {\bibinfo {author} {\bibnamefont {{Safranova M.S. \it{et
  al.}}}},\ }\href {https://doi.org/10.1103/RevModPhys.90.025008} {\bibfield
  {journal} {\bibinfo  {journal} {Rev. Mod. Phys.}\ }\textbf {\bibinfo {volume}
  {90}} (\bibinfo {year} {2018})}\BibitemShut {NoStop}%
\bibitem [{\citenamefont {{Davoudiasl H. \it{et al.}}}(2012)}]{Dav12}%
  \BibitemOpen
  \bibfield  {author} {\bibinfo {author} {\bibnamefont {{Davoudiasl H. \it{et
  al.}}}},\ }\href {https://doi.org/10.1103/PhysRevD.85.115019} {\bibfield
  {journal} {\bibinfo  {journal} {Phys. Rev. D}\ }\textbf {\bibinfo {volume}
  {85}} (\bibinfo {year} {2012})}\BibitemShut {NoStop}%
\bibitem [{\citenamefont {{Langacker P.}}(2009)}]{Lan09}%
  \BibitemOpen
  \bibfield  {author} {\bibinfo {author} {\bibnamefont {{Langacker P.}}},\
  }\href {https://doi.org/10.1103/RevModPhys.81.1199} {\bibfield  {journal}
  {\bibinfo  {journal} {Rev. Mod. Phys.}\ }\textbf {\bibinfo {volume} {81}}
  (\bibinfo {year} {2009})}\BibitemShut {NoStop}%
\bibitem [{\citenamefont {{Wood C. \it{et al.}}}(1997)}]{Woo97}%
  \BibitemOpen
  \bibfield  {author} {\bibinfo {author} {\bibnamefont {{Wood C. \it{et
  al.}}}},\ }\href {https://doi.org/10.1126/science.275.5307.1759} {\bibfield
  {journal} {\bibinfo  {journal} {Science}\ }\textbf {\bibinfo {volume} {275}}
  (\bibinfo {year} {1997})}\BibitemShut {NoStop}%
\bibitem [{\citenamefont {{The Jefferson Lab PVDIS
  Collaboration.}}(2014)}]{PVDIS}%
  \BibitemOpen
  \bibfield  {author} {\bibinfo {author} {\bibnamefont {{The Jefferson Lab
  PVDIS Collaboration.}}},\ }\href {https://doi.org/10.1038/nature12964}
  {\bibfield  {journal} {\bibinfo  {journal} {Nature}\ }\textbf {\bibinfo
  {volume} {506}},\ \bibinfo {pages} {67} (\bibinfo {year} {2014})}\BibitemShut
  {NoStop}%
\bibitem [{\citenamefont {{The Jefferson Lab Qweak
  Collaboration}}(2020)}]{QWEAK}%
  \BibitemOpen
  \bibfield  {author} {\bibinfo {author} {\bibnamefont {{The Jefferson Lab
  Qweak Collaboration}}},\ }\href {https://doi.org/10.1103/PhysRevC.101.055503}
  {\bibfield  {journal} {\bibinfo  {journal} {Phys. Rev. C}\ }\textbf {\bibinfo
  {volume} {101}},\ \bibinfo {pages} {055503} (\bibinfo {year}
  {2020})}\BibitemShut {NoStop}%
\bibitem [{\citenamefont {{Desplanques B., Donoghue J.F., \& Holstein
  B.R.}}(1980)}]{Des80}%
  \BibitemOpen
  \bibfield  {author} {\bibinfo {author} {\bibnamefont {{Desplanques B.,
  Donoghue J.F., \& Holstein B.R.}}},\ }\href
  {https://doi.org/10.1016/0003-4916(80)90217-1} {\bibfield  {journal}
  {\bibinfo  {journal} {Ann. Phys.}\ }\textbf {\bibinfo {volume} {124}},\
  \bibinfo {pages} {449} (\bibinfo {year} {1980})}\BibitemShut {NoStop}%
\bibitem [{\citenamefont {{Adelberger E.G. \& Haxton W.C.}}(1985)}]{Ade85}%
  \BibitemOpen
  \bibfield  {author} {\bibinfo {author} {\bibnamefont {{Adelberger E.G. \&
  Haxton W.C.}}},\ }\href {https://doi.org/10.1146/annurev.ns.35.120185.002441}
  {\bibfield  {journal} {\bibinfo  {journal} {Ann. Rev. Nucl. Part. Sc.}\
  }\textbf {\bibinfo {volume} {35}},\ \bibinfo {pages} {501} (\bibinfo {year}
  {1985})}\BibitemShut {NoStop}%
\bibitem [{\citenamefont {{Haxton W.C. \& Holstein B.R.}}(2013)}]{Hax13}%
  \BibitemOpen
  \bibfield  {author} {\bibinfo {author} {\bibnamefont {{Haxton W.C. \&
  Holstein B.R.}}},\ }\href
  {https://doi.org/https://doi.org/10.1016/j.ppnp.2013.03.009} {\bibfield
  {journal} {\bibinfo  {journal} {Progress in Particle and Nuclear Physics}\
  }\textbf {\bibinfo {volume} {71}},\ \bibinfo {pages} {185} (\bibinfo {year}
  {2013})}\BibitemShut {NoStop}%
\bibitem [{\citenamefont {{Gardner S., Haxton W.C. \& Holstein
  B.R.}}(2017)}]{Gar17}%
  \BibitemOpen
  \bibfield  {author} {\bibinfo {author} {\bibnamefont {{Gardner S., Haxton
  W.C. \& Holstein B.R.}}},\ }\href
  {https://doi.org/10.1146/annurev-nucl-041917-033231} {\bibfield  {journal}
  {\bibinfo  {journal} {Ann. Rev. Nucl. Part. Sc.}\ }\textbf {\bibinfo {volume}
  {67}},\ \bibinfo {pages} {69} (\bibinfo {year} {2017})}\BibitemShut {NoStop}%
\bibitem [{\citenamefont {{de Vries J. \it{et al.}}}(2020)}]{deV20}%
  \BibitemOpen
  \bibfield  {author} {\bibinfo {author} {\bibnamefont {{de Vries J. \it{et
  al.}}}},\ }\href {https://doi.org/10.3389/fphy.2020.00218} {\bibfield
  {journal} {\bibinfo  {journal} {Front. Phys.}\ }\textbf {\bibinfo {volume}
  {8}} (\bibinfo {year} {2020})}\BibitemShut {NoStop}%
\bibitem [{\citenamefont {{Gardner S. \& Muralidhara G.}}(2023)}]{Gar23}%
  \BibitemOpen
  \bibfield  {author} {\bibinfo {author} {\bibnamefont {{Gardner S. \&
  Muralidhara G.}}},\ }\href {https://doi.org/10.1103/PhysRevC.107.055501}
  {\bibfield  {journal} {\bibinfo  {journal} {Phys. Rev. C}\ }\textbf {\bibinfo
  {volume} {107}},\ \bibinfo {pages} {055501} (\bibinfo {year}
  {2023})}\BibitemShut {NoStop}%
\bibitem [{\citenamefont {{Ramsey-Musolf M.J. \& Page S.A.}}(2006)}]{Ram06}%
  \BibitemOpen
  \bibfield  {author} {\bibinfo {author} {\bibnamefont {{Ramsey-Musolf M.J. \&
  Page S.A.}}},\ }\href {https://doi.org/10.1146/annurev.nucl.54.070103.181255}
  {\bibfield  {journal} {\bibinfo  {journal} {Annu. Rev. Nucl. Part. Sci.}\
  }\textbf {\bibinfo {volume} {56}},\ \bibinfo {pages} {1} (\bibinfo {year}
  {2006})}\BibitemShut {NoStop}%
\bibitem [{\citenamefont {{Holstein B.R.}}(2007)}]{Hol07}%
  \BibitemOpen
  \bibfield  {author} {\bibinfo {author} {\bibnamefont {{Holstein B.R.}}},\
  }\href {https://doi.org/10.1140/epja/i2006-10430-0} {\bibfield  {journal}
  {\bibinfo  {journal} {Eur. Phys. J. A}\ }\textbf {\bibinfo {volume} {32}},\
  \bibinfo {pages} {505} (\bibinfo {year} {2007})}\BibitemShut {NoStop}%
\bibitem [{\citenamefont {{ACME Collaboration}}(2018)}]{ACM18}%
  \BibitemOpen
  \bibfield  {author} {\bibinfo {author} {\bibnamefont {{ACME
  Collaboration}}},\ }\href {https://doi.org/10.1038/s41586-018-0599-8}
  {\bibfield  {journal} {\bibinfo  {journal} {Nature}\ }\textbf {\bibinfo
  {volume} {562}} (\bibinfo {year} {2018})}\BibitemShut {NoStop}%
\bibitem [{\citenamefont {{Cairncross W.B. \it{et al.}}}(2017)}]{Cai17}%
  \BibitemOpen
  \bibfield  {author} {\bibinfo {author} {\bibnamefont {{Cairncross W.B. \it{et
  al.}}}},\ }\href {https://doi.org/10.1103/PhysRevLett.119.153001} {\bibfield
  {journal} {\bibinfo  {journal} {Phys. Rev. Lett.}\ }\textbf {\bibinfo
  {volume} {119}} (\bibinfo {year} {2017})}\BibitemShut {NoStop}%
\bibitem [{\citenamefont {{Roussy T.S. \it{et al.}}}(2023)}]{Rou22}%
  \BibitemOpen
  \bibfield  {author} {\bibinfo {author} {\bibnamefont {{Roussy T.S. \it{et
  al.}}}},\ }\href {https://doi.org/10.1126/science.adg4084} {\bibfield
  {journal} {\bibinfo  {journal} {Science}\ }\textbf {\bibinfo {volume}
  {381}},\ \bibinfo {pages} {46} (\bibinfo {year} {2023})}\BibitemShut
  {NoStop}%
\bibitem [{\citenamefont {{Garcia Ruiz R.F. \it{et al.}}}(2020)}]{Gar20}%
  \BibitemOpen
  \bibfield  {author} {\bibinfo {author} {\bibnamefont {{Garcia Ruiz R.F.
  \it{et al.}}}},\ }\href {https://doi.org/10.1038/s41586-020-2299-4}
  {\bibfield  {journal} {\bibinfo  {journal} {Nature}\ }\textbf {\bibinfo
  {volume} {581}} (\bibinfo {year} {2020})}\BibitemShut {NoStop}%
\bibitem [{\citenamefont {{Sushkov O.P. \& Flambaum V.V.}}(1978)}]{Sus78}%
  \BibitemOpen
  \bibfield  {author} {\bibinfo {author} {\bibnamefont {{Sushkov O.P. \&
  Flambaum V.V.}}},\ }\href {http://jetp.ras.ru/cgi-bin/dn/e_048_04_0608.pdf}
  {\bibfield  {journal} {\bibinfo  {journal} {Zh. Eksp. Teor. Fiz}\ }\textbf
  {\bibinfo {volume} {75}},\ \bibinfo {pages} {1208} (\bibinfo {year}
  {1978})}\BibitemShut {NoStop}%
\bibitem [{\citenamefont {{DeMille D.P. \it{et al.}}}(2008)}]{DeM08}%
  \BibitemOpen
  \bibfield  {author} {\bibinfo {author} {\bibnamefont {{DeMille D.P. \it{et
  al.}}}},\ }\href {http://doi.org/10.1103/PhysRevLett.100.023003} {\bibfield
  {journal} {\bibinfo  {journal} {Phys. Rev. Lett.}\ }\textbf {\bibinfo
  {volume} {100}} (\bibinfo {year} {2008})}\BibitemShut {NoStop}%
\bibitem [{\citenamefont {{Flambaum V.V. \& Khriplovich I.B.}}(1985)}]{Fla85}%
  \BibitemOpen
  \bibfield  {author} {\bibinfo {author} {\bibnamefont {{Flambaum V.V. \&
  Khriplovich I.B.}}},\ }\href {https://doi.org/10.1016/0375-9601(85)90756-X}
  {\bibfield  {journal} {\bibinfo  {journal} {Phys. Lett. A}\ }\textbf
  {\bibinfo {volume} {110}} (\bibinfo {year} {1985})}\BibitemShut {NoStop}%
\bibitem [{\citenamefont {{Kozlov M.G. \& Labzowsky L.N.}}(1995)}]{Koz95}%
  \BibitemOpen
  \bibfield  {author} {\bibinfo {author} {\bibnamefont {{Kozlov M.G. \&
  Labzowsky L.N.}}},\ }\href {https://doi.org/10.1088/0953-4075/28/10/008}
  {\bibfield  {journal} {\bibinfo  {journal} {J. Phys. B}\ }\textbf {\bibinfo
  {volume} {28}} (\bibinfo {year} {1995})}\BibitemShut {NoStop}%
\bibitem [{\citenamefont {{Altuntas E. \it{et
  al.}}}(2018{\natexlab{a}})}]{Alt18}%
  \BibitemOpen
  \bibfield  {author} {\bibinfo {author} {\bibnamefont {{Altuntas E. \it{et
  al.}}}},\ }\href {https://doi.org/10.1103/PhysRevLett.120.142501} {\bibfield
  {journal} {\bibinfo  {journal} {Phys. Rev. Lett.}\ }\textbf {\bibinfo
  {volume} {120}} (\bibinfo {year} {2018}{\natexlab{a}})}\BibitemShut {NoStop}%
\bibitem [{\citenamefont {{Udrescu S.M. \it{et al.}}}(2021)}]{Udr21}%
  \BibitemOpen
  \bibfield  {author} {\bibinfo {author} {\bibnamefont {{Udrescu S.M. \it{et
  al.}}}},\ }\href {https://doi.org/10.1103/PhysRevLett.127.033001} {\bibfield
  {journal} {\bibinfo  {journal} {Phys. Rev. Lett.}\ }\textbf {\bibinfo
  {volume} {127}} (\bibinfo {year} {2021})}\BibitemShut {NoStop}%
\bibitem [{\citenamefont {{Arrowsmith-Kron G. \it{et
  al.}}}(2023)}]{RadMol2023}%
  \BibitemOpen
  \bibfield  {author} {\bibinfo {author} {\bibnamefont {{Arrowsmith-Kron G.
  \it{et al.}}}},\ }\href {https://arxiv.org/abs/2302.02165} {\bibfield
  {journal} {\bibinfo  {journal} {arXiv}\ }\textbf {\bibinfo {volume}
  {2302.02165}} (\bibinfo {year} {2023})}\BibitemShut {NoStop}%
\bibitem [{\citenamefont {{Flambaum V.V. \& Khriplovich I.B.}}(1980)}]{Fla80}%
  \BibitemOpen
  \bibfield  {author} {\bibinfo {author} {\bibnamefont {{Flambaum V.V. \&
  Khriplovich I.B.}}},\ }\href {https://www.osti.gov/biblio/6280930} {\bibfield
   {journal} {\bibinfo  {journal} {Sov. Phys. JETP}\ }\textbf {\bibinfo
  {volume} {52}},\ \bibinfo {pages} {835} (\bibinfo {year} {1980})}\BibitemShut
  {NoStop}%
\bibitem [{\citenamefont {{Flambaum V.V., Khriplovich I.B. \& Sushkov
  O.P.}}(1984)}]{Fla84}%
  \BibitemOpen
  \bibfield  {author} {\bibinfo {author} {\bibnamefont {{Flambaum V.V.,
  Khriplovich I.B. \& Sushkov O.P.}}},\ }\href
  {https://doi.org/10.1016/0370-2693(84)90140-0} {\bibfield  {journal}
  {\bibinfo  {journal} {Phys. Lett. B}\ }\textbf {\bibinfo {volume} {146}},\
  \bibinfo {pages} {367} (\bibinfo {year} {1984})}\BibitemShut {NoStop}%
\bibitem [{\citenamefont {{Dzuba V.A. \& Flambaum V.V.}}(2000)}]{Dzu00}%
  \BibitemOpen
  \bibfield  {author} {\bibinfo {author} {\bibnamefont {{Dzuba V.A. \& Flambaum
  V.V.}}},\ }\href {https://doi.org/10.1103/PhysRevA.62.052101} {\bibfield
  {journal} {\bibinfo  {journal} {Phys. Rev. A}\ }\textbf {\bibinfo {volume}
  {62}},\ \bibinfo {pages} {052101} (\bibinfo {year} {2000})}\BibitemShut
  {NoStop}%
\bibitem [{\citenamefont {{Johnson W.R., Safronova M.S. \& Safronova
  U.I.}}(2003)}]{Joh03}%
  \BibitemOpen
  \bibfield  {author} {\bibinfo {author} {\bibnamefont {{Johnson W.R.,
  Safronova M.S. \& Safronova U.I.}}},\ }\href
  {https://doi.org/10.1103/PhysRevA.67.062106} {\bibfield  {journal} {\bibinfo
  {journal} {Phys. Rev. A}\ }\textbf {\bibinfo {volume} {67}},\ \bibinfo
  {pages} {062106} (\bibinfo {year} {2003})}\BibitemShut {NoStop}%
\bibitem [{\citenamefont {{Dzuba V.A., Flambaum V.V. \& Stadnik
  Y.V.}}(2017)}]{Dzu17}%
  \BibitemOpen
  \bibfield  {author} {\bibinfo {author} {\bibnamefont {{Dzuba V.A., Flambaum
  V.V. \& Stadnik Y.V.}}},\ }\href
  {https://doi.org/10.1103/PhysRevLett.119.223201} {\bibfield  {journal}
  {\bibinfo  {journal} {Phys. Rev. Lett.}\ }\textbf {\bibinfo {volume} {119}},\
  \bibinfo {pages} {223201} (\bibinfo {year} {2017})}\BibitemShut {NoStop}%
\bibitem [{\citenamefont {{Stadnik Y.V. \& Flambaum V.V.}}(2015)}]{Sta15}%
  \BibitemOpen
  \bibfield  {author} {\bibinfo {author} {\bibnamefont {{Stadnik Y.V. \&
  Flambaum V.V.}}},\ }\href {https://doi.org/10.1140/epjc/s10052-015-3326-8}
  {\bibfield  {journal} {\bibinfo  {journal} {Eur. Phys. J. C}\ }\textbf
  {\bibinfo {volume} {75}},\ \bibinfo {pages} {110} (\bibinfo {year}
  {2015})}\BibitemShut {NoStop}%
\bibitem [{\citenamefont {{Stadnik Y.V. \& Flambaum V.V.}}(2014)}]{Sta14}%
  \BibitemOpen
  \bibfield  {author} {\bibinfo {author} {\bibnamefont {{Stadnik Y.V. \&
  Flambaum V.V.}}},\ }\href {https://doi.org/10.1103/PhysRevD.89.043522}
  {\bibfield  {journal} {\bibinfo  {journal} {Phys. Rev. D}\ }\textbf {\bibinfo
  {volume} {89}},\ \bibinfo {pages} {043522} (\bibinfo {year}
  {2014})}\BibitemShut {NoStop}%
\bibitem [{\citenamefont {{Stollenwerk P.R. \it{et al.}}}(2016)}]{Sto16}%
  \BibitemOpen
  \bibfield  {author} {\bibinfo {author} {\bibnamefont {{Stollenwerk P.R.
  \it{et al.}}}},\ }\href {https://doi.org/10.1016/j.jms.2016.09.012}
  {\bibfield  {journal} {\bibinfo  {journal} {J. Mol. Spec.}\ }\textbf
  {\bibinfo {volume} {332}} (\bibinfo {year} {2016})}\BibitemShut {NoStop}%
\bibitem [{\citenamefont {{Stollenwerk P.R. \it{et al.}}}(2020)}]{Sto20}%
  \BibitemOpen
  \bibfield  {author} {\bibinfo {author} {\bibnamefont {{Stollenwerk P.R.
  \it{et al.}}}},\ }\href {https://doi.org/10.1103/PhysRevLett.125.113201}
  {\bibfield  {journal} {\bibinfo  {journal} {Phys. Rev. Lett.}\ }\textbf
  {\bibinfo {volume} {125}} (\bibinfo {year} {2020})}\BibitemShut {NoStop}%
\bibitem [{\citenamefont {{Nguyen J.H.V. \& Odom B.}}(2011)}]{Ngu11}%
  \BibitemOpen
  \bibfield  {author} {\bibinfo {author} {\bibnamefont {{Nguyen J.H.V. \& Odom
  B.}}},\ }\href {https://doi.org/10.1103/PhysRevA.83.053404} {\bibfield
  {journal} {\bibinfo  {journal} {Phys. Rev. A}\ }\textbf {\bibinfo {volume}
  {83}} (\bibinfo {year} {2011})}\BibitemShut {NoStop}%
\bibitem [{\citenamefont {{Kozlov M.G., Labzovski L.N. \& Mitrushchenko
  A.O.}}(1991)}]{Koz91}%
  \BibitemOpen
  \bibfield  {author} {\bibinfo {author} {\bibnamefont {{Kozlov M.G., Labzovski
  L.N. \& Mitrushchenko A.O.}}},\ }\href
  {http://qchem.pnpi.spb.ru/kozlov/My_papers/KLM_JETP.pdf} {\bibfield
  {journal} {\bibinfo  {journal} {Zh. Éksp. Teor. Fiz.}\ }\textbf {\bibinfo
  {volume} {100}} (\bibinfo {year} {1991})}\BibitemShut {NoStop}%
\bibitem [{\citenamefont {{Zhu G.Z. \it{et al.}}}(2022)}]{Zhu22}%
  \BibitemOpen
  \bibfield  {author} {\bibinfo {author} {\bibnamefont {{Zhu G.Z. \it{et
  al.}}}},\ }\href {https://doi.org/10.1016/j.jms.2022.111582} {\bibfield
  {journal} {\bibinfo  {journal} {J. Mol. Spec.}\ }\textbf {\bibinfo {volume}
  {384}},\ \bibinfo {pages} {111582} (\bibinfo {year} {2022})}\BibitemShut
  {NoStop}%
\bibitem [{\citenamefont {{Hao Y. \it{et al.}}}(2018)}]{Hao18}%
  \BibitemOpen
  \bibfield  {author} {\bibinfo {author} {\bibnamefont {{Hao Y. \it{et
  al.}}}},\ }\href {https://doi.org/10.1103/PhysRevA.98.032510} {\bibfield
  {journal} {\bibinfo  {journal} {Phys. Rev. A}\ }\textbf {\bibinfo {volume}
  {98}} (\bibinfo {year} {2018})}\BibitemShut {NoStop}%
\bibitem [{\citenamefont {{Knight Jr. L.B. \it{et
  al.}}}(1985)}]{knight1985generation}%
  \BibitemOpen
  \bibfield  {author} {\bibinfo {author} {\bibnamefont {{Knight Jr. L.B. \it{et
  al.}}}},\ }\href {https://doi.org/10.1021/ja00296a005} {\bibfield  {journal}
  {\bibinfo  {journal} {J. Am. Chem. Soc.}\ }\textbf {\bibinfo {volume}
  {107}},\ \bibinfo {pages} {2857} (\bibinfo {year} {1985})}\BibitemShut
  {NoStop}%
\bibitem [{\citenamefont {{Zhu G.-Z. \it{et al.}}}(2022)}]{zhu2022high}%
  \BibitemOpen
  \bibfield  {author} {\bibinfo {author} {\bibnamefont {{Zhu G.-Z. \it{et
  al.}}}},\ }\href {https://doi.org/10.1016/j.jms.2022.111582} {\bibfield
  {journal} {\bibinfo  {journal} {J. Mol. Spectrosc.}\ }\textbf {\bibinfo
  {volume} {384}},\ \bibinfo {pages} {111582} (\bibinfo {year}
  {2022})}\BibitemShut {NoStop}%
\bibitem [{\citenamefont {{Filianin P. \it{et al.}}}(2021)}]{Fil21}%
  \BibitemOpen
  \bibfield  {author} {\bibinfo {author} {\bibnamefont {{Filianin P. \it{et
  al.}}}},\ }\href {https://doi.org/10.1103/PhysRevLett.127.072502} {\bibfield
  {journal} {\bibinfo  {journal} {Phys. Rev. Lett.}\ }\textbf {\bibinfo
  {volume} {127}} (\bibinfo {year} {2021})}\BibitemShut {NoStop}%
\bibitem [{\citenamefont {{Sailer T. \it{et al.}}}(2022)}]{Sai22}%
  \BibitemOpen
  \bibfield  {author} {\bibinfo {author} {\bibnamefont {{Sailer T. \it{et
  al.}}}},\ }\href {https://doi.org/10.1038/s41586-022-04807-w} {\bibfield
  {journal} {\bibinfo  {journal} {Nature}\ }\textbf {\bibinfo {volume} {606}}
  (\bibinfo {year} {2022})}\BibitemShut {NoStop}%
\bibitem [{\citenamefont {{Mougeot M. \it{et al.}}}(2021)}]{Mou21}%
  \BibitemOpen
  \bibfield  {author} {\bibinfo {author} {\bibnamefont {{Mougeot M. \it{et
  al.}}}},\ }\href {https://doi.org/10.1038/s41567-021-01326-9} {\bibfield
  {journal} {\bibinfo  {journal} {Nature Phys.}\ }\textbf {\bibinfo {volume}
  {17}} (\bibinfo {year} {2021})}\BibitemShut {NoStop}%
\bibitem [{\citenamefont {{Borchert M.J. \it{et al.}}}(2022)}]{Bor22}%
  \BibitemOpen
  \bibfield  {author} {\bibinfo {author} {\bibnamefont {{Borchert M.J. \it{et
  al.}}}},\ }\href {https://doi.org/10.1038/s41586-021-04203-w} {\bibfield
  {journal} {\bibinfo  {journal} {Nature}\ }\textbf {\bibinfo {volume} {601}}
  (\bibinfo {year} {2022})}\BibitemShut {NoStop}%
\bibitem [{\citenamefont {{Brown L.S. \& Gabrielse G.}}(1986)}]{Bro86}%
  \BibitemOpen
  \bibfield  {author} {\bibinfo {author} {\bibnamefont {{Brown L.S. \&
  Gabrielse G.}}},\ }\href {https://doi.org/10.1103/RevModPhys.58.233}
  {\bibfield  {journal} {\bibinfo  {journal} {Rev. Mod. Phys.}\ }\textbf
  {\bibinfo {volume} {58}} (\bibinfo {year} {1986})}\BibitemShut {NoStop}%
\bibitem [{\citenamefont {{Martinez F. \it{et al.}}}(2014)}]{Mar14}%
  \BibitemOpen
  \bibfield  {author} {\bibinfo {author} {\bibnamefont {{Martinez F. \it{et
  al.}}}},\ }\href {https://doi.org/10.1016/j.ijms.2013.12.018} {\bibfield
  {journal} {\bibinfo  {journal} {Int. J. Mass Spec.}\ }\textbf {\bibinfo
  {volume} {365}},\ \bibinfo {pages} {266} (\bibinfo {year}
  {2014})}\BibitemShut {NoStop}%
\bibitem [{\citenamefont {{Cahn S.B. \it{et al.}}}(2014)}]{Cah14}%
  \BibitemOpen
  \bibfield  {author} {\bibinfo {author} {\bibnamefont {{Cahn S.B. \it{et
  al.}}}},\ }\href {https://doi.org/10.1103/PhysRevLett.112.163002} {\bibfield
  {journal} {\bibinfo  {journal} {Phys. Rev. Lett.}\ }\textbf {\bibinfo
  {volume} {112}},\ \bibinfo {pages} {163002} (\bibinfo {year}
  {2014})}\BibitemShut {NoStop}%
\bibitem [{\citenamefont {{Cornell E.A. \it{et al.}}}(1990)}]{Cor90}%
  \BibitemOpen
  \bibfield  {author} {\bibinfo {author} {\bibnamefont {{Cornell E.A. \it{et
  al.}}}},\ }\href {https://doi.org/10.1103/PhysRevA.41.312} {\bibfield
  {journal} {\bibinfo  {journal} {Phys. Rev. A}\ }\textbf {\bibinfo {volume}
  {41}} (\bibinfo {year} {1990})}\BibitemShut {NoStop}%
\bibitem [{\citenamefont {{Jensen M., Hasegawa T. \& Bollinger
  J.}}(2004)}]{Jen04}%
  \BibitemOpen
  \bibfield  {author} {\bibinfo {author} {\bibnamefont {{Jensen M., Hasegawa T.
  \& Bollinger J.}}},\ }\href {https://doi.org/10.1103/PhysRevA.70.033401}
  {\bibfield  {journal} {\bibinfo  {journal} {J. Phys. Rev. A}\ }\textbf
  {\bibinfo {volume} {70}} (\bibinfo {year} {2004})}\BibitemShut {NoStop}%
\bibitem [{\citenamefont {{Comisarow M.B. \& Marshall A.G.}}(1974)}]{Com74}%
  \BibitemOpen
  \bibfield  {author} {\bibinfo {author} {\bibnamefont {{Comisarow M.B. \&
  Marshall A.G.}}},\ }\href {https://doi.org/10.1016/0009-2614(74)89137-2}
  {\bibfield  {journal} {\bibinfo  {journal} {Chem. Phys. Lett.}\ }\textbf
  {\bibinfo {volume} {25}} (\bibinfo {year} {1974})}\BibitemShut {NoStop}%
\bibitem [{\citenamefont {{Schüssler R.X. \it{et al.}}}(2020)}]{Sch20}%
  \BibitemOpen
  \bibfield  {author} {\bibinfo {author} {\bibnamefont {{Schüssler R.X. \it{et
  al.}}}},\ }\href {https://doi.org/10.1038/s41586-020-2221-0} {\bibfield
  {journal} {\bibinfo  {journal} {Nature}\ }\textbf {\bibinfo {volume} {581}}
  (\bibinfo {year} {2020})}\BibitemShut {NoStop}%
\bibitem [{\citenamefont {{Marr A.J., Flores M. \& Steimle
  T.C.}}(1996)}]{Mar96}%
  \BibitemOpen
  \bibfield  {author} {\bibinfo {author} {\bibnamefont {{Marr A.J., Flores M.
  \& Steimle T.C.}}},\ }\href {https://doi.org/10.1063/1.471573} {\bibfield
  {journal} {\bibinfo  {journal} {J. Chem. Phys.}\ }\textbf {\bibinfo {volume}
  {104}} (\bibinfo {year} {1996})}\BibitemShut {NoStop}%
\bibitem [{\citenamefont {{Stollenwerk P.R., Antonov I.O. \& Odom
  B.C.}}(2019)}]{Sto19}%
  \BibitemOpen
  \bibfield  {author} {\bibinfo {author} {\bibnamefont {{Stollenwerk P.R.,
  Antonov I.O. \& Odom B.C.}}},\ }\href
  {https://doi.org/10.1016/j.jms.2018.11.008} {\bibfield  {journal} {\bibinfo
  {journal} {J. Mol. Spec.}\ }\textbf {\bibinfo {volume} {355}} (\bibinfo
  {year} {2019})}\BibitemShut {NoStop}%
\bibitem [{\citenamefont {{Tong X., Winney A. \& Willitsch S.}}(2010)}]{Ton10}%
  \BibitemOpen
  \bibfield  {author} {\bibinfo {author} {\bibnamefont {{Tong X., Winney A. \&
  Willitsch S.}}},\ }\href {https://doi.org/10.1103/PhysRevLett.105.143001}
  {\bibfield  {journal} {\bibinfo  {journal} {Phys. Rev. Lett.}\ }\textbf
  {\bibinfo {volume} {105}} (\bibinfo {year} {2010})}\BibitemShut {NoStop}%
\bibitem [{\citenamefont {{Andresen G.B. \it{et al.}}}(2010)}]{And10}%
  \BibitemOpen
  \bibfield  {author} {\bibinfo {author} {\bibnamefont {{Andresen G.B. \it{et
  al.}}}},\ }\href {http://doi.org/10.1103/PhysRevLett.105.013003} {\bibfield
  {journal} {\bibinfo  {journal} {Phys. Rev. Lett.}\ }\textbf {\bibinfo
  {volume} {105}} (\bibinfo {year} {2010})}\BibitemShut {NoStop}%
\bibitem [{\citenamefont {{Sturm S. \it{et al.}}}(2019)}]{Stu19}%
  \BibitemOpen
  \bibfield  {author} {\bibinfo {author} {\bibnamefont {{Sturm S. \it{et
  al.}}}},\ }\href {https://doi.org/10.1140/epjst/e2018-800225-2} {\bibfield
  {journal} {\bibinfo  {journal} {Eur. Phys. J. Spec. Top.}\ }\textbf {\bibinfo
  {volume} {227}} (\bibinfo {year} {2019})}\BibitemShut {NoStop}%
\bibitem [{\citenamefont {{Altuntas E. \it{et
  al.}}}(2018{\natexlab{b}})}]{Alt18b}%
  \BibitemOpen
  \bibfield  {author} {\bibinfo {author} {\bibnamefont {{Altuntas E. \it{et
  al.}}}},\ }\href {https://doi.org/10.1103/PhysRevA.97.042101} {\bibfield
  {journal} {\bibinfo  {journal} {Phys. Rev. A}\ }\textbf {\bibinfo {volume}
  {97}},\ \bibinfo {pages} {042101} (\bibinfo {year}
  {2018}{\natexlab{b}})}\BibitemShut {NoStop}%
\bibitem [{\citenamefont {{Chmaisani W. \& Elmoussaoui S.}}(2021)}]{Chm21}%
  \BibitemOpen
  \bibfield  {author} {\bibinfo {author} {\bibnamefont {{Chmaisani W. \&
  Elmoussaoui S.}}},\ }\href {https://doi.org/10.1039/D0CP05575A} {\bibfield
  {journal} {\bibinfo  {journal} {Phys. Chem. Chem. Phys.}\ }\textbf {\bibinfo
  {volume} {23}},\ \bibinfo {pages} {1718} (\bibinfo {year}
  {2021})}\BibitemShut {NoStop}%
\bibitem [{\citenamefont {{Takeda Y. \it{et al.}}}(2022)}]{Tak22}%
  \BibitemOpen
  \bibfield  {author} {\bibinfo {author} {\bibnamefont {{Takeda Y. \it{et
  al.}}}},\ }\href {https://doi.org/10.1088/1361-6668/ac5645} {\bibfield
  {journal} {\bibinfo  {journal} {Supercond. Sci. Technol.}\ }\textbf {\bibinfo
  {volume} {35}} (\bibinfo {year} {2022})}\BibitemShut {NoStop}%
\bibitem [{\citenamefont {{Britton J.W. \it{et al.}}}(2016)}]{Bri16}%
  \BibitemOpen
  \bibfield  {author} {\bibinfo {author} {\bibnamefont {{Britton J.W. \it{et
  al.}}}},\ }\href {https://doi.org/10.1103/PhysRevA.93.062511} {\bibfield
  {journal} {\bibinfo  {journal} {Phys. Rev. A}\ }\textbf {\bibinfo {volume}
  {93}},\ \bibinfo {pages} {062511} (\bibinfo {year} {2016})}\BibitemShut
  {NoStop}%
\bibitem [{\citenamefont {{Droese C. \it{et al.}}}(2011)}]{Dro11}%
  \BibitemOpen
  \bibfield  {author} {\bibinfo {author} {\bibnamefont {{Droese C. \it{et
  al.}}}},\ }\href {https://doi.org/10.1016/j.nima.2010.12.176} {\bibfield
  {journal} {\bibinfo  {journal} {Nucl. Instr. Meth. Phys. Res. A}\ }\textbf
  {\bibinfo {volume} {632}} (\bibinfo {year} {2011})}\BibitemShut {NoStop}%
\bibitem [{\citenamefont {{Dyall K.G.}}(2006)}]{Dya06}%
  \BibitemOpen
  \bibfield  {author} {\bibinfo {author} {\bibnamefont {{Dyall K.G.}}},\ }\href
  {https://doi.org/10.1007/s00214-006-0126-0} {\bibfield  {journal} {\bibinfo
  {journal} {Theor. Chem. Acc.}\ }\textbf {\bibinfo {volume} {115}} (\bibinfo
  {year} {2006})}\BibitemShut {NoStop}%
\bibitem [{\citenamefont {{Dyall K.G.}}(2016)}]{Dya16}%
  \BibitemOpen
  \bibfield  {author} {\bibinfo {author} {\bibnamefont {{Dyall K.G.}}},\ }\href
  {https://doi.org/10.1007/s00214-016-1884-y} {\bibfield  {journal} {\bibinfo
  {journal} {Theor. Chem. Acc.}\ }\textbf {\bibinfo {volume} {135}} (\bibinfo
  {year} {2016})}\BibitemShut {NoStop}%
\bibitem [{\citenamefont {{Gomes A.S.P. \it{et al.}}}(2019)}]{Gom19}%
  \BibitemOpen
  \bibfield  {author} {\bibinfo {author} {\bibnamefont {{Gomes A.S.P. \it{et
  al.}}}},\ }\href {https://doi.org/10.5281/zenodo.3572669} {\bibfield
  {journal} {\bibinfo  {journal} {Zenodo}\ } (\bibinfo {year}
  {2019})}\BibitemShut {NoStop}%
\bibitem [{\citenamefont {{Saue, T. \it{et al.}}}(2020)}]{Sau20}%
  \BibitemOpen
  \bibfield  {author} {\bibinfo {author} {\bibnamefont {{Saue, T. \it{et
  al.}}}},\ }\href {https://doi.org/10.1063/5.0004844} {\bibfield  {journal}
  {\bibinfo  {journal} {J. Chem. Phys.}\ }\textbf {\bibinfo {volume} {152}}
  (\bibinfo {year} {2020})}\BibitemShut {NoStop}%
\bibitem [{\citenamefont {{Dyall K.G.}}(2002)}]{Dya02}%
  \BibitemOpen
  \bibfield  {author} {\bibinfo {author} {\bibnamefont {{Dyall K.G.}}},\ }\href
  {http://doi.org/10.1007/s00214-002-0388-0} {\bibfield  {journal} {\bibinfo
  {journal} {Theor. Chem. Acc.}\ }\textbf {\bibinfo {volume} {108}} (\bibinfo
  {year} {2002})}\BibitemShut {NoStop}%
\bibitem [{\citenamefont {{Dyall K.G.}}(2007)}]{Dya07}%
  \BibitemOpen
  \bibfield  {author} {\bibinfo {author} {\bibnamefont {{Dyall K.G.}}},\ }\href
  {https://doi.org/10.1007/s00214-006-0175-4} {\bibfield  {journal} {\bibinfo
  {journal} {Theor. Chem. Acc.}\ }\textbf {\bibinfo {volume} {117}} (\bibinfo
  {year} {2007})}\BibitemShut {NoStop}%
\bibitem [{\citenamefont {{Ilias M. \& Saue T.}}(2007)}]{Ili07}%
  \BibitemOpen
  \bibfield  {author} {\bibinfo {author} {\bibnamefont {{Ilias M. \& Saue
  T.}}},\ }\href {https://doi.org/10.1063/1.2436882} {\bibfield  {journal}
  {\bibinfo  {journal} {J. Chem. Phys.}\ }\textbf {\bibinfo {volume} {126}}
  (\bibinfo {year} {2007})}\BibitemShut {NoStop}%
\bibitem [{\citenamefont {{Saue. T.}}(2011)}]{Sau11}%
  \BibitemOpen
  \bibfield  {author} {\bibinfo {author} {\bibnamefont {{Saue. T.}}},\ }\href
  {https://doi.org/10.1002/cphc.201100682} {\bibfield  {journal} {\bibinfo
  {journal} {Chem. Phys. Chem.}\ }\textbf {\bibinfo {volume} {12}} (\bibinfo
  {year} {2011})}\BibitemShut {NoStop}%
\bibitem [{\citenamefont {{Pototschnig J.V \it{et al.}}}(2021)}]{Pot21}%
  \BibitemOpen
  \bibfield  {author} {\bibinfo {author} {\bibnamefont {{Pototschnig J.V \it{et
  al.}}}},\ }\href {https://doi.org/10.1021/acs.jctc.1c00260} {\bibfield
  {journal} {\bibinfo  {journal} {J. Chem. Theo. Comp.}\ }\textbf {\bibinfo
  {volume} {17}} (\bibinfo {year} {2021})}\BibitemShut {NoStop}%
\bibitem [{\citenamefont {{Lagerqvist A., Renhorn I. \& Elander
  N.}}(1973)}]{Lag73}%
  \BibitemOpen
  \bibfield  {author} {\bibinfo {author} {\bibnamefont {{Lagerqvist A., Renhorn
  I. \& Elander N.}}},\ }\href {https://doi.org/10.1016/0022-2852(73)90043-X}
  {\bibfield  {journal} {\bibinfo  {journal} {J. Mol. Spec.}\ }\textbf
  {\bibinfo {volume} {46}} (\bibinfo {year} {1973})}\BibitemShut {NoStop}%
\bibitem [{\citenamefont {{Kurth T. \it{et al.}}}(2016)}]{Kur16}%
  \BibitemOpen
  \bibfield  {author} {\bibinfo {author} {\bibnamefont {{Kurth T. \it{et
  al.}}}},\ }\href {https://doi.org/10.22323/1.251.0329} {\bibfield  {journal}
  {\bibinfo  {journal} {Proc. Sci.}\ }\textbf {\bibinfo {volume} {251}}
  (\bibinfo {year} {2016})}\BibitemShut {NoStop}%
\bibitem [{\citenamefont {{Davoudi Z. \it{et al.}}}(2021)}]{Dav21}%
  \BibitemOpen
  \bibfield  {author} {\bibinfo {author} {\bibnamefont {{Davoudi Z. \it{et
  al.}}}},\ }\href {https://doi.org/10.1016/j.physrep.2020.10.004} {\bibfield
  {journal} {\bibinfo  {journal} {Phys. Rep.}\ }\textbf {\bibinfo {volume}
  {700}} (\bibinfo {year} {2021})}\BibitemShut {NoStop}%
\bibitem [{\citenamefont {{Hao Y. \it{et al.}}}(2020)}]{Hao20}%
  \BibitemOpen
  \bibfield  {author} {\bibinfo {author} {\bibnamefont {{Hao Y. \it{et
  al.}}}},\ }\href {https://doi.org/10.1103/PhysRevA.102.052828} {\bibfield
  {journal} {\bibinfo  {journal} {Phys. Rev. A}\ }\textbf {\bibinfo {volume}
  {102}} (\bibinfo {year} {2020})}\BibitemShut {NoStop}%
\bibitem [{\citenamefont {{Norrgard E.B. \it{et al.}}}(2019)}]{Nor19}%
  \BibitemOpen
  \bibfield  {author} {\bibinfo {author} {\bibnamefont {{Norrgard E.B. \it{et
  al.}}}},\ }\href {https://doi.org/10.1038/s42005-019-0181-1} {\bibfield
  {journal} {\bibinfo  {journal} {Comm. Phys.}\ }\textbf {\bibinfo {volume}
  {2}},\ \bibinfo {pages} {77} (\bibinfo {year} {2019})}\BibitemShut {NoStop}%
\bibitem [{\citenamefont {{Bohman M. \it{et al.}}}(2021)}]{Boh21}%
  \BibitemOpen
  \bibfield  {author} {\bibinfo {author} {\bibnamefont {{Bohman M. \it{et
  al.}}}},\ }\href {https://doi.org/10.1038/s41586-021-03784-w} {\bibfield
  {journal} {\bibinfo  {journal} {Nature}\ }\textbf {\bibinfo {volume} {596}}
  (\bibinfo {year} {2021})}\BibitemShut {NoStop}%
\bibitem [{\citenamefont {{Will C. \it{et al.}}}(2022)}]{Wil22}%
  \BibitemOpen
  \bibfield  {author} {\bibinfo {author} {\bibnamefont {{Will C. \it{et
  al.}}}},\ }\href {https://doi.org/10.1088/1367-2630/ac55b3} {\bibfield
  {journal} {\bibinfo  {journal} {New J. Phys.}\ }\textbf {\bibinfo {volume}
  {24}} (\bibinfo {year} {2022})}\BibitemShut {NoStop}%
\bibitem [{\citenamefont {{Will C. }}(2023)}]{Wil23}%
  \BibitemOpen
  \bibfield  {author} {\bibinfo {author} {\bibnamefont {{Will C. }}},\ }\href
  {https://doi.org/10.48550/arXiv.2310.10208} {\bibfield  {journal} {\bibinfo
  {journal} {In Preprint at arXiv}\ ,\ \bibinfo {pages} {2310.10208}} (\bibinfo
  {year} {2023})}\BibitemShut {NoStop}%
\bibitem [{\citenamefont {{Itano W.M. \& Wineland D.J.}}(1982)}]{Ita82}%
  \BibitemOpen
  \bibfield  {author} {\bibinfo {author} {\bibnamefont {{Itano W.M. \& Wineland
  D.J.}}},\ }\href {https://doi.org/10.1103/PhysRevA.25.35} {\bibfield
  {journal} {\bibinfo  {journal} {Phys. Rev. A}\ }\textbf {\bibinfo {volume}
  {25}} (\bibinfo {year} {1982})}\BibitemShut {NoStop}%
\bibitem [{\citenamefont {{Torrisi S.B. \it{et al.}}}(2016)}]{Tor16}%
  \BibitemOpen
  \bibfield  {author} {\bibinfo {author} {\bibnamefont {{Torrisi S.B. \it{et
  al.}}}},\ }\href {https://doi.org/10.1103/PhysRevA.93.043421} {\bibfield
  {journal} {\bibinfo  {journal} {Phys. Rev. A}\ }\textbf {\bibinfo {volume}
  {93}} (\bibinfo {year} {2016})}\BibitemShut {NoStop}%
\bibitem [{\citenamefont {{Borschevsky A.}}(2012)}]{Bor12}%
  \BibitemOpen
  \bibfield  {author} {\bibinfo {author} {\bibnamefont {{Borschevsky A.}}},\
  }\href {http://doi.org/10.1103/PhysRevA.86.050501} {\bibfield  {journal}
  {\bibinfo  {journal} {Phys. Rev. A}\ }\textbf {\bibinfo {volume} {86}}
  (\bibinfo {year} {2012})}\BibitemShut {NoStop}%
\bibitem [{\citenamefont {{Budker D., Kimball D.F. \& DeMille
  D.P.}}(2004)}]{budker2004atomic}%
  \BibitemOpen
  \bibfield  {author} {\bibinfo {author} {\bibnamefont {{Budker D., Kimball
  D.F. \& DeMille D.P.}}},\ }\href
  {https://books.google.com/books?id=GW6pclAk-JcC} {\bibfield  {journal}
  {\bibinfo  {journal} {Oxford Press}\ } (\bibinfo {year} {2004})}\BibitemShut
  {NoStop}%
\bibitem [{\citenamefont {{Chattopadhyaya S. \it{et al.}}}(2003)}]{Cha03}%
  \BibitemOpen
  \bibfield  {author} {\bibinfo {author} {\bibnamefont {{Chattopadhyaya S.
  \it{et al.}}}},\ }\href {https://doi.org/10.1016/j.theochem.2003.08.007}
  {\bibfield  {journal} {\bibinfo  {journal} {J. Mol. Struc.}\ }\textbf
  {\bibinfo {volume} {639}},\ \bibinfo {pages} {177} (\bibinfo {year}
  {2003})}\BibitemShut {NoStop}%
\bibitem [{\citenamefont {{Arfken G.}}(1985)}]{arfken1985spherical}%
  \BibitemOpen
  \bibfield  {author} {\bibinfo {author} {\bibnamefont {{Arfken G.}}},\
  }\href@noop {} {\bibfield  {journal} {\bibinfo  {journal} {Academic Press
  Orlando}\ }\textbf {\bibinfo {volume} {1}},\ \bibinfo {pages} {680} (\bibinfo
  {year} {1985})}\BibitemShut {NoStop}%
\end{thebibliography}%




\newpage
\onecolumngrid
\titleformat{\section}[hang]{\large\bfseries}{\thesection}{0.5em}{}[]
\titlespacing*{\section}{10pt}{*4}{5pt}

\newcommand{\PRLsep}{\noindent\makebox[\linewidth]{\resizebox{0.75\linewidth}{1.75pt}{$\bullet$}}\bigskip}

\appendix
\begin{center}
    \vspace{0.5cm}
    \PRLsep
    \vspace{0.5cm}
    \textbf{\Large Supplemental Material}
\end{center}
\section{| Effects of Time-Varying Electric Fields}\label{sec:rad-E-field}

A time-varying electric field can shift the energy levels of the ion through the AC Stark shift, thus changing the effective splitting between the two levels of interest. Therefore, any uncertainty in the magnitude of such a field will manifest itself as an uncertainty in the level splitting and, consequently, in the extracted parity violation signal. The AC Stark shift due to the electric dipole moment coupling between two levels of opposite parity separated by $\omega_0$ is given by \cite{budker2004atomic}:
\begin{equation}
    \Delta E = \frac{\varOmega_{\rm{R}}^2}{2}\frac{\omega_0}{\omega_0^2-\omega_{\rm{ext}}^2}
    \label{general_AC_shift_formula}
\end{equation}
where $\omega_{\rm{ext}}$ and $\varOmega_{\rm{R}}$ are the frequency and Rabi frequency of the external electric field, respectively. 

We are interested in the levels that come close to degeneracy in the presence of the applied magnetic field. These levels have $N=0$ and $N=1$. For this analysis, only the contribution of states with $N=0, 1, 2$ has been considered when calculating the AC Stark shift. Levels with $N > 2$ will have a significantly smaller effect as they contribute only at higher orders in perturbation theory and are located further away in frequency from the states of interest. The location of the considered states in frequency space was obtained by diagonalizing the effective molecular Hamiltonian in the presence of the magnetic field, using the decoupled basis $\ket{N,m_N}\ket{S,m_S}\ket{I,m_I}$:
\begin{equation}
    \begin{split}
        H &= H_{\rm{eff}} + H_{\rm{mag}} \\
        H_{\rm{eff}} &= B\boldsymbol{N}^2+\gamma\boldsymbol{N}\cdot\boldsymbol{S}+b\boldsymbol{I}\cdot\boldsymbol{S}+c\left(\boldsymbol{I}\cdot\boldsymbol{n}\right)\left(\boldsymbol{S}\cdot\boldsymbol{n}\right)\\
        H_{\rm{mag}} &= -g_\perp\mu_B\boldsymbol{S}\cdot\boldsymbol{B}-\left(g_\parallel-g_\perp\right)\mu_B\left(\boldsymbol{S}\cdot\boldsymbol{n}\right)\left(\boldsymbol{B}\cdot\boldsymbol{n}\right)-g_I\mu_N\boldsymbol{I}\cdot\boldsymbol{B}\\
    \end{split}
\end{equation}
The various spectroscopic constants were taken from \cite{knight1985generation,zhu2022high}.

In our case, the main time-varying electric fields felt by the ion in its rest frame are the axial one, produced by the axial electric field of the trap and the radial ones due to the quadrupolar shape of the trap's electric field and the induced $v \times B$ electric field due to the ion's magnetron and modified cyclotron motions. Any radial field will be experienced as a rotating field in the ion's rest frame. The total external electric field can therefore be written as:

\begin{equation}
    \begin{split}
        \overrightarrow{E}(t) = E_z\hat{z}\cos\left(\omega_zt+\phi_z\right) + E_{\perp}\left(\hat{x}\cos\omega_\perp t+\hat{y}\sin\omega_\perp t\right),
    \end{split}
\end{equation}
where $E_z$ and $E_\perp$ are the magnitudes of the axial and transverse electric field (among the radial electric fields, the one due to the ion's cyclotron motion dominates), $\omega_z$ and $\omega_\perp$ are the axial and cyclotron frequencies and $\phi_z$ is a phase difference between the axial and radial fields. To calculate the Rabi frequency, $\varOmega_{\rm{R}}$, the operator:

\begin{equation}
    \begin{split}
        \overrightarrow{d}\cdot\overrightarrow{E} = d E_z n_z \cos \left(\omega_z t + \phi_z\right) + \frac{dE_\perp}{2}\left(n_+ e^{-i\omega_\perp t} + n_- e^{i\omega_\perp t}\right)
    \end{split}
    \label{formula_for_AC_shift}
\end{equation}
needs to be evaluated between different rotational/hyperfine levels of interest. The internuclear axis operator $\hat{n}$ in the lab frame is given by:
\begin{equation}
    \begin{split}
        \hat{n} = \sin\theta\cos\phi\hat{x}+\sin\theta\sin\phi\hat{y}+\cos\theta\hat{z},
    \end{split}
\end{equation}
which can be rewritten in spherical harmonics form using the following:
\begin{equation}
    \begin{split}
        n_z &= \cos\theta=2\sqrt{\frac{\pi}{3}}Y_1^0(\theta,\phi), \\
        n_+ &= n_x+in_y = \sin\theta e^{i\phi} = -2\sqrt{\frac{2\pi}{3}}Y_1^1(\theta,\phi), \\
        n_- &= n_x-in_y = \sin\theta e^{-i\phi}=2\sqrt{\frac{2\pi}{3}}Y_1^{-1}(\theta,\phi).
    \end{split}
\end{equation}
To calculate $\varOmega_{\rm{R}}$, the dipole moment of $^{29}$SiO$^+$ in its ground electronic state, $d$, is needed. This has not been measured experimentally yet, so this analysis will use the theoretically predicted value of $d = 4.147\,$D \cite{Cha03}. The calculated AC Stark shifts scale as $\propto d^2$, so the obtained values can be adjusted accordingly later once $d$ is measured.  

The matrix elements required for the calculation of $\varOmega_{\rm{R}}$ are of the form $\braket{N,m_N|dE|N',m_N'}$, where:

\begin{equation}
    \begin{split}
        \ket{N,m_N} &= Y_N^{m_N}(\theta,\phi), \\
        \bra{N,m_N} &= \overline{Y_N^{m_N}}(\theta,\phi) = (-1)^{m_N}Y_N^{-m_N}(\theta,\phi).
    \end{split}
\end{equation}
The required integrals can be evaluated using \cite{arfken1985spherical}:
\begin{equation}
    \int_0^{2\pi}\int_0^\pi Y_{l_1}^{m_1}Y_{l_2}^{m_2}Y_{l_3}^{m_3}\sin\theta d\theta d\phi = \sqrt{\frac{(2l_1+1)(2l_2+1)(2l_3+1)}{4\pi}} \begin{pmatrix}
l_1 & l_2 & l_3\\
0 & 0 & 0
\end{pmatrix}\begin{pmatrix}
l_1 & l_2 & l_3\\
m_1 & m_2 & m_3
\end{pmatrix}
\end{equation}
The $\overrightarrow{d}\cdot\overrightarrow{E}$ operator does not act in the electron or nuclear spin space. Therefore, the effective electric dipole moment between levels with different values of $m_S$ or $m_I$ is given by $d_{\rm{eff}} \sim \eta d$, with $\eta \sim (b,c,\gamma)/B \ll 1$ \cite{Alt18}.

In our experiment, different avoided level crossing will remove various systematic effects \cite{Alt18}. For any two such levels brought close to degeneracy, we calculate the matrix elements of the operator in Eq. \ref{formula_for_AC_shift} between each of these two levels and all the other levels considered in this analysis. The result is plugged into Eq. \ref{general_AC_shift_formula}, together with the previously calculated values for $\omega_0$, to get the AC Stark shift of the two levels of interest. A similar value for the AC Stark shift is obtained for all the pairs of investigated avoided crossing levels. The uncertainty on the shift due to the radial field AC Stark shift, $\Delta^{\mathrm{rad}}_{\rm{AC}}/2\pi \propto \alpha_r E_\perp^2 \propto r_c^2$, is dominated by the uncertainty on the radius of the cyclotron motion of the ion, $r_{\rm{c}}$, and at out temperature ($T \approx 1$ K) it amounts to $\updelta\Delta^{\mathrm{rad}}_{\rm{AC}}/2\pi \approx 10\,$Hz. The uncertainty due to the axial AC Stark shift, $\Delta^{\mathrm{axial}}_{\rm{AC}}/2\pi \propto \alpha_z E_z^2$, comes mainly from the uncertainty on $E_z$ due to the thermal electric field and at our temperature it gives $\updelta\Delta^{\mathrm{axial}}_{\rm{AC}}/2\pi \approx 20 \ \mathrm{Hz}$.

\section{| Asymmetry Analytical Formula}\label{sec:analyt-formula} 

The Hamiltonian of the two levels of opposite parity brought close to degeneracy, in the presence of the parity violation interaction and a time-varying electric field is:
\begin{equation}H_{\pm}=
    \begin{pmatrix}
0 & iW+\varOmega_{\rm{R}}\sin\left(\omega_{\rm{ext}} t\right) &\\
-iW+\varOmega_{\rm{R}}\sin\left(\omega_{\rm{ext}} t\right) & \varDelta
\end{pmatrix},
\end{equation}
where $iW$ is the imaginary parity violating matrix element and $\omega_{\rm{ext}}$ and $\varOmega_{\rm{R}}$ are the oscillating frequency and the Rabi frequency of the electric field, respectively.

In our experiment, we populate the positive parity state and measure the population transfer to the negative parity state after an interrogation time $t_{\rm{x}}$. If this population transfer is small ($\lesssim 10\%$), as it is expected in the case of $^{29}$SiO$^+$, we can get insight into the evolution of the system using first-order time-dependent perturbation theory. In this case, the population transfer after a time $t$ is given by:
\begin{equation}
    \begin{split}
        |c_-(t)|^2 &= \left|\frac{2W}{\varDelta}e^{-i\frac{\varDelta t}{2}}\sin{\left(\frac{\varDelta t}{2}\right)}+i\frac{\varOmega_{\rm{R}}}{\omega_{\rm{ext}}}\left(\cos{(\omega_{\rm{ext}} t)e^{-i\varDelta t}}-1\right)\right|^2, \\
        |c_-(t)|^2 &= (r_1+r_2)^2+(r_1+r_2\cos{(\omega_{\rm{ext}} t)})^2 - 2(r_1+r_2)(r_1+r_2\cos{(\omega_{\rm{ext}} t)})\cos{(\varDelta t)}
    \end{split}
\end{equation}
with $r_1=\frac{W}{\varDelta}$ and $r_2=\frac{\varOmega_{\rm{R}}}{\omega_{\rm{ext}}}$. The asymmetry is then given by:
\begin{equation}
    A_{\rm{PV}} = \frac{2r_1r_2(1+\cos{(\omega_{\rm{ext}} t)})(1-\cos{(\varDelta t)})}{2r_1^2(1-\cos{(\varDelta t)})+r_2^2[1+\cos^2{(\omega_{\rm{ext}} t)}-2\cos{(\omega t)\cos{(\varDelta t)}}]}.
\end{equation}
If the measurement is performed at $t\approx \frac{2\pi N}{\omega_{\rm{ext}}}\approx\frac{\pi}{\varDelta}$ \cite{DeM08, Alt18} for integer $N$, we end up with:
\begin{equation}
     |c_-(t)|^2 = 4\left(r_1+r_2\right)^2
\end{equation}
and
\begin{equation}
    A_{\rm{PV}} = \frac{2r_1r_2}{r_1^2+r_2^2} = \frac{2\frac{W}{\varDelta}\frac{\varOmega_{\rm{R}}}{\omega_{\rm{ext}}}}{\left(\frac{W}{\varDelta}\right)^2+\left(\frac{\varOmega_{\rm{R}}}{\omega_{\rm{ext}}}\right)^2}.
\end{equation}

The initial and final "kicks" applied to the ion in order to produce its oscillatory motion will change the population transfer predicted by the formulas above by a small amount, given that the length of each "kick" (few $\mu$s) is expected to be many orders of magnitude smaller than the actual measurement time (few ms). The actual change in population can be easily calculated once the shape and duration of the "kicks" are known for a given measurement.

\end{document}